\documentclass[a4paper,12pt]{article}

\usepackage[utf8]{inputenc}
\usepackage{amsmath,amsfonts,bbm,amsthm,mathtools}
\usepackage{xcolor,setspace,wasysym}
\usepackage{multicol,enumitem}
\usepackage{url}
\usepackage{graphicx,caption}
\usepackage{authblk}
\usepackage{natbib}

\usepackage{lineno}
\modulolinenumbers[5]

\RequirePackage[colorlinks=true,urlcolor=blue!50!black, citecolor=blue!50!black,linkcolor=blue!50!black]{hyperref}

\makeatletter
\def\namedlabel#1#2{\begingroup
   \def\@currentlabel{#2}%
   \label{#1}\endgroup
}
\makeatother

\DeclareMathOperator*{\argmin}{argmin}
\DeclareMathOperator*{\argmax}{argmax}
\DeclareMathOperator*{\argsup}{argsup}

\DeclareMathOperator*{\sign}{sgn}

\newcommand{\R}{\mathbb{R}}
\def\1{\mbox{1\hspace{-.35em}1}}
\theoremstyle{plain}

\newtheorem{proposition}{Proposition}
\newtheorem{definition}{Definition}
\newtheorem{rem}{Remark}

\title{Deepest Voting: a new way of electing}

\author[1]{Jean-Baptiste Aubin}
\author[1]{Irène Gannaz}
\author[1]{Samuela Leoni}
\author[2]{Antoine Rolland} 

\date{December 2021}

\affil[1]{Univ Lyon, INSA Lyon, UJM, UCBL, ECL, ICJ, UMR5208, \protect\\ 69621 Villeurbanne, France\protect\\~}
\affil[2]{ERIC EA 3083, Universit\'e de Lyon, Université Lumi\`ere Lyon 2, \protect\\ 5 Pierre Mendès France, 69596 Bron Cedex, France}

\begin{document}
%%%%%%%%%%%%%%%%%%%%%%%%%%%%%%%%%

\setlength{\parindent}{0pt}
\setlength{\parskip}{\baselineskip}

\maketitle

\onehalfspacing

\begin{abstract}
This article aims to present a unified framework for grading-based voting processes. The idea is to represent the grades of each voter on $d$ candidates as a point in $\R^d$ and to define the winner of the vote using the deepest point of the scatter plot. The deepest point is obtained by the maximization of a depth function. Universality, unanimity, and neutrality properties are proved to be satisfied. Monotonicity and Independence to Irrelevant Alternatives are also studied. It is shown that usual voting processes correspond to specific choices of depth functions. Finally, some basic paradoxes are explored for these voting processes.
\end{abstract}


\newpage

\mathtoolsset{showonlyrefs} 

\section{Introduction }\label{sec:objectifs}

Balinski and Laraki \citep{Balinski2007,Balinski2014,Balinski2020} have developed the ``theory of measuring, electing and ranking''. The authors show that voting procedures, based on evaluations rather than on rankings, satisfy valuable properties. They propose a voting rule, the majority judgment, based on the medians of the evaluations and prove that it does not fall in the scope of Arrow’s impossibility theorem \citep{Balinski2007}. This property emerges from the information contained in each vote. This framework, based on the grading of the candidates by the voters, gives more nuanced information than a ranking-based setting. The grading model has encountered much interest in the last decades, and alternatives to majority judgment are, for example, approval voting \citep{Brams} and range voting \citep{Smith}.

This article aims to present a unified framework for grading-based voting processes, study some of their properties, and extend the scope of voting processes. This family of social decision functions is based on the statistical notions of depth functions and their related deepest points. Let us consider in the following that we have $n$ voters and $d$ candidates. Each voter gives a grade to each candidate. Each voter can then be assimilated to a point in $\mathbb{R}^d$,  whose coordinates are the grades for each candidate. The set of all voters' grades can, hence, be seen as a scatter plot. The key idea  is to consider the \emph{most central} voter in this scatter plot. According to their grades, this innermost (possibly imaginary) voter can be seen as the most representative of all voters. Hence, their preferences should meet a large consensus among the others voters. Therefore, the presented social decision function returns simply the candidate who has the maximum grade of this innermost voter. We refer to any voting process based on the use of such notion of the \emph{most central} voter as a ``deepest voting'' process.

The definition of the deepest point of a scatter plot is a well-known research topic in statistics. It was initiated in 1975 by Tukey, who  defined a multivariate median of a given multivariate data cloud \citep{Tukey}. The idea is to introduce a depth function, which associates to each point a value, so that the depth function is maximal at the innermost point of the scatter plot and minimal for outliers. Many notions of data depth have been introduced since then, proposing alternative definitions of the deepest point. See \emph{e.g.} the monographs of \cite{ZuoSerfling} and \cite{Mosler}. Depth functions vary regarding their computability and robustness and their sensitivity to reflect the shapes of the data.  Our claim is that it can bring an interesting viewpoint for voting processes. The properties of the depths functions are linked with the properties of the associated voting process and some new voting procedures can be proposed,  based on different notions of depth functions.

The paper is organized as follows. Section~\ref{sec:voting} recalls the definition of \cite{Balinski2007}'s grading model.
 In Section~\ref{sec:depth}, we recall the statistical notion of depth function which will help us to determine the innermost voter, knowing that there are many ways of choosing the ``center" of a scatter plot. Each way of choosing it corresponds to a new member of the family of social decision functions. We next give the definition of the deepest voting process. We show in Section \ref{sec:conditions} that main usual conditions on voting processes --non-dictatorship, universality, unanimity-- are satisfied by classical depth functions. We also characterize monotonicity and independence to irrelevant alternatives with respect to the behavior of the depth functions. We establish that some usual depth functions such as halfspace or projection, do not lead to monotone voting procedures, which seems a main drawback. Finally, we study some interesting properties for a given family of depth functions in Section~\ref{sec:wLp}. This family includes, to our knowledge, all grading-based voting processes, namely, majority judgment, range voting and approval voting. We show that voting processes of this family may suffer from 
 {Condorcet winner, Condorcet looser, reinforcement and no-show paradoxes}. We provide a discussion about the pertinence of viewing these properties as paradoxes in the grading-based context. All the proofs are given in the Appendix.

\section{Voting framework}

\label{sec:voting}

\subsection{Voting process}

A voting process can be seen as a mathematical function and subsequently voting processes have been widely studied from a mathematical point of view since the early works of Borda and Condorcet at the end of the XVIII$^{th}$ century (see \cite{Felsenthal} for a review). This research field is known as ``social choice theory''. The well-known impossibility theorems of \cite{Arrow} or \cite{Gibbard} and \cite{Satterthwaite} demonstrate that no voting process can simultaneously satisfy a minimal set of desirable properties. Therefore the choice of a voting process appears as a matter of compromise between pro and cons arguments. One can refer to  \cite{Felsenthal} for a complete review of the properties/paradoxes satisfied by the most popular voting processes.

Another way is to change the paradigm of the voting situation so that voters do not only rank the candidates but also rate them.

\subsection{Grade modeling}\label{sec:grades}

Grading candidates rather than ranking them allows using a specific voting process out of the framework of Arrow's theorem. If voters are supposed to grade the candidates, then the voting process consists of finding the best candidate considering all grades. 
Approval voting \citep{Brams} is the simplest example of such a grading-based voting process, where grades are 0 or 1 and the chosen candidate is the one with the greatest number of 1. The majority judgment \citep{Balinski2007} is another example of a grading-based voting process, using discrete or continuous grades. We propose in the following to formalize the use of grades in the voting process as in  \cite{Balinski2007}.

Consider that we have $n$ voters and $d$ candidates.
Suppose that each voter $v_j$, $j=1,\dots,n$ grades each candidate $c_i$, $i=1,\dots,d$. We denote $\Phi(i,j)$ the corresponding grade. Let $\Lambda$ be the set of possible grades $\Phi(i,j)$. As pointed out by \cite{Balinski2007}, the set of grades $\Lambda$ needs to be strictly ordered but might be finite or an interval of the real numbers. We suppose without loss of generality that $\Lambda\subseteq[0,1]$.
The grading is summarized in a $d \times n$ grading matrix $\Phi=\{\Phi(i,j), i=1,\dots,d, j=1,\dots,n\}\in\Lambda^{d\times n}$. Any collection 
of $d$ grades is called a \emph{profile} and, in particular, every column $\Phi(.,j)$ constitutes a profile.

We distinguish three cases depending on the nature of the set $\Lambda$:
\begin{description}
\item[Binary set] $\Lambda=\{0, 1\}$.  Approval voting is a voting process based on a binary set of grades. First-past-the-post system may correspond to the case where $\Phi(i,j)=1$ for one and only one candidate $c_i$ for each voter $v_j$, whereas approval voting gives no constraint on the number of 0 or 1 by a voter.

\item[Discrete set] (e.g. $\Lambda=\{0, 1/N, 2/N \dots,1\}$, with $N>1$). This case includes \emph{e.g.} grading with finite words from bad to excellent, letters from E to A, etc, where a numerical ranking is applied. A usual example in everyday life is the evaluation process of a product or of a service, where each consumer is asked to put a mark between 1 and 5.

\item[Continuous set.] The set $\Lambda$ is a real interval. Let $\Lambda=[0, 1]$ without loss of generality. In practice, generalization of classical voting procedure to continuous sets $\Lambda$ can be processed by putting an horizontal segment in front of each candidate's name and asking a voter to put a mark on this segment indicating their level of accordance with the candidate.
\end{description}

If formally a procedure is proposed with a continuous set $\Lambda$, in practice, a discretization is necessary and the number of observations is always finite. Hence,  $\Lambda$ can always be treated as a discrete set. An advantage of our framework is that the fact that $\Lambda$ is discrete or continuous does not have any impact on the decision procedure.

\paragraph{Example} \label{sec:example}
Table \ref{tab:examcontinudiscretbinaire} illustrates with an example the different types of grading. Consider 9 voters and 2 candidates. The continuous grading allows each voter to give any grade between 0 and 1 to any candidate. The discrete grading (here on 11 levels, \emph{i.e.} $N=10$) can be seen as a rounding of continuous grading. Binary grading corresponds to a rounding of continuous grading or setting the maximal grade to 1 and others to 0 for majority voting. 

\begin{table}[!ht]
\centering
\begin{tabular}{llccccccccc} \hline
voter  & & $v_1$ &$v_2$ &$v_3$ &$v_4$ &$v_5$ &$v_6$ &$v_7$ &$v_8$ &$v_9$ \\
\hline
\textrm{continuous} & $c_1$ & 0.14 & 0.38 & 0.34 & 0.43 & 0.45 & 0.61 & 0.84 & 0.69 & 0.95 \\
\textrm{grading} & $c_2$ & 0.43  & 0.14 & 0.68 & 0.80 & 0.64 & 0.75 & 0.66 & 0.48 & 0.16 \\
 \hline
\textrm{discrete} & $c_1$ & 0.1 & 0.5 & 0.3 & 0.4 & 0.4 & 0.6 & 0.8 & 0.7 & 0.9 \\
\textrm{grading} & $c_2$ & 0.4 & 0.1 & 0.7 & 0.8 & 0.6 & 0.7 & 0.7 & 0.5 & 0.2\\
\hline
\textrm{binary} & $c_1$ &  0 & 0 & 0 & 0 & 0 & 1 & 1 & 1 & 1  \\
 \textrm{grading} & $c_2$ & 0 & 0 & 1 & 1 & 1 & 1 & 1 & 0 & 0 \\
\hline
\end{tabular}
\caption{Example of grades given by 9 voters on 2 candidates with continuous, discrete or binary scales. \label{tab:examcontinudiscretbinaire}}
\end{table}

\subsection{Grading-based voting process} \label{sec:conditionsgrading}

A grading-based voting process can then be seen as a function $G$, called \emph{method of grading} in~\cite{Balinski2007}, assigning a profile to any matrix $\Phi$. The function $G$ is defined from $\Lambda^{d \times n}$ with values in the subsets of $[0,1]^d$. Note that the set of possible grades of the profiles given by function $G$ may differs from $\Lambda$ since one may associate for example the mean of initial grades and obtain a value which possibly does not belong to $\Lambda$. 

\cite{Balinski2007} propose some conditions (which they call \emph{axioms}) that a \emph{method of grading} should satisfy:
\begin{description}[font=\bf]
\item[{Neutrality}\namedlabel{C1}{\emph{Neutrality}}]  %$G$ is \emph{neutral}: 
$G$ gives the same result by permuting the rows of $\Phi$ (\emph{i.e.} by permuting the candidates).
\item[{Universality}\namedlabel{C2}{\emph{Universality}}]  %$G$ is \emph{anonymous}: 
$G$ gives the same result by permuting the columns of $\Phi$ (\emph{i.e.} by permuting the voters).
\item[{Unanimity}\namedlabel{C3}{\emph{Unanimity}}] 
if a candidate is given an identical grade $\alpha$ by every voter, then $G$ assigns him the grade $\alpha$. 
\item[{Monotonicity}\namedlabel{C4}{\emph{Monotonicity}}]  
if $\Phi = \widetilde\Phi$ except that one or more voters give higher grades to candidate $c_i$ in $ \Phi $ than in $ \widetilde\Phi $, then $G(\Phi)(c_i)$ is higher than $G(\widetilde\Phi)(c_i)$.
\item[{IIA}\namedlabel{C5}{\emph{IIA}}] (Independence of Irrelevant Alternatives) 
 if the grades assigned by the voters to a candidate $c_i$ in two profiles $\Phi$ and $\widetilde\Phi$ are the same, then $G(\Phi)(c_i)=G(\widetilde\Phi)(c_i)$.
\end{description}
\vspace{0.2cm}

These conditions are similar to the ones used in Arrow's theorem \citep{Arrow}. Therefore, in the ranking-based model, it is impossible to find a voting process that satisfies all these conditions. This has lead to the proposal of a grading-based voting process by \cite{Balinski2007} to overcome this drawback. 

The simplest way to compute a grading-based voting process is to aggregate the grades given by the voters on each candidate independently. Note that such a process satisfies \ref{C5}. 
In such a case, we introduce the aggregation function $g:\Lambda^{n} \to [0,1]$. The function $g$ summarizes the $n$ grades received by a candidate $c_i$ (the row i of $\Phi$) in a unique grade. One has $G(\Phi)(c_i)=g(\Phi(i,\cdot))$. Many aggregation functions are available to sum up $n$ grades into a unique one (see \cite{Beliakov} or \cite{Grabisch09} for a review of aggregation functions). Some of them have been studied within a specific framework of grading-based voting process, taking into account the fact that votes often take place in a political context.

\cite{Balinski2007} propose the \emph{majority-grade} voting as an aggregation function. For a given candidate $c_i$,  let rank the $n$ grades $\{\Phi(i,j), j=1,\dots,n\}$ as $r_1 \leq r_2 \leq \hdots \leq r_n$. The majority-grade voting, denoted $g^{maj}$, is defined as follows:
\begin{equation}\label{eqn:mj}
g^{maj}(r_1,\hdots,r_n)=\begin{cases}r_{(n+1)/2} & \textrm{ if $n$ is odd,}\\ r_{(n+2)/2} & \textrm{ if $n$ is even.}
\end{cases}
\end{equation}
$g^{maj}(r_1,\hdots,r_n)$ can be interpreted as a median of $(r_1,\hdots,r_n)$  A \emph{majority-ranking} can be deduced from the majority-grade, evidently noting that a candidate receiving a higher majority-grade than another will be ranked higher. Note that tie-breaking rules have been proposed for example in \cite{Balinski2020} or \cite{Fabre}.

\cite{Smith} studies another example of aggregation function: the \emph{range voting} $g^{rv}$, defined by 
\begin{equation*}
\forall (r_1,\hdots,r_n) \in \Lambda^n, g^{rv}(r_1,\hdots,r_n)=\displaystyle \frac{1}{n} \sum_{j=1}^n r_j,
\end{equation*}
with similar notations with the above. $g^{rv}(r_1,\hdots,r_n)$ is the mean of the grades $r_1,\hdots,r_n$. Observe that approval voting uses the same function on a binary set $\Lambda=\{0;1\}$.

Grading-based voting processes take advantage of more information than a mere preference order on the candidates than ranking-based voting processes since they introduce the intensity of the relative preferences. This supplementary amount of information permits overcoming classical impossibility theorems.
We propose in the following a unified framework for a grading-based voting process that enlightens the specificity of both majority judgment and range voting within a wide range of new voting processes.

\section{Deepest voting}

\label{sec:depth}

Our statement is that depth functions facilitate the consideration of classical grading models and continuous grading models in a uniform way. It expands to numerous voting processes. We first recall what a depth function is, and we present, on a second hand, how it applies in voting processes.

\subsection{What is a depth function?}

Quoting \cite{ZuoSerfling}: ``Associated with a given distribution $F$ on $\mathbb{R}^d$, a depth function is designed to provide a $F$-based center-outward ordering [...] of points $x$ in $\mathbb{R}^d$. High depth corresponds to \textit{centrality}, low depth to \textit{outlyingness}''. In other words, a depth function takes high (positive) values at the ``middle" of a scatter plot and vanishes out of it.

Denote by $\mathcal{F}$ the class of distributions on the Borel sets of $\mathbb{R}^d$ and $F_X$ the distribution of a given random vector $X$. We define a depth function as follows.

\begin{definition} \label{def:depth} Let the mapping $D:\mathbb{R}^d \times \mathcal{F} \to \mathbb{R}$ be bounded, nonnegative and satisfying:
\begin{enumerate}[label=(P\arabic*)]
\item Let $X=(X_1,\dots,X_d)$ be a random vector in $\R^d$, $x\in\R^d$, and $\sigma$ a permutation on $\{1,\dots,d\}$. Let $X_\sigma=(X_{\sigma(1)},\dots,X_{\sigma(d)})$ and $x_\sigma=(x_{\sigma(1)},\dots,x_{\sigma(d)})$. Then $D(x_\sigma,F_{X_\sigma})=D(x; F_X)$.
\label{ass:depth0}

\item For all $a\in\R$, $b\in\R^d$, for any random vector $X \in \mathbb{R}^d$, $\argmax_{x\in\R^d} D(a\, x+b;F_{a\,X+b})= \argmax_{x\in\R^d} D(x;F_{X})$.
\label{ass:depth1}

\item For a distribution $F \in \mathcal{F}$ having a uniquely defined ``center'' $\theta$ (\emph{e.g.} the point of ``symmetry''), $D(\theta;F)=\sup_{x \in \mathbb{R}^d} D(x;F)$.
\label{ass:depth2} 

\item For any $F \in \mathcal{F}$, $D(\cdot;F)$ is quasi-concave. That is, if $\theta\in\argmax_{x\in\R^d} D(x;F)$, then $D(x;F) \leq D(\theta + \lambda (x-\theta); F)$ for any $0 \leq \lambda \leq 1$. \label{ass:depth3}

\item $D(x;F) \rightarrow 0$ as $ \|x\| \rightarrow \infty$ for each  $F \in \mathcal{F}$. \label{ass:depth4}

\item Let $F \in \mathcal{F}$ be a distribution on $\R^d$ with marginal distributions $F_1,\dots, F_d$. Suppose that for $i\in\{1,\dots,d\}$, $F_i$ has support containing a unique point $\{\alpha\}$. Then for all $x^*\in\argsup_{x \in \mathbb{R}^d} D(x;F)$, the $i^\text{th}$ coordinate of $x^*$ is $x^*_i=\alpha$. \label{ass:depth5}

\end{enumerate}
Then $D(\cdot;\cdot)$ is called a {\bf statistical depth function}.

\end{definition}

Assumptions \ref{ass:depth0} and \ref{ass:depth1} are often replaced by a stronger assumption, which is:
\begin{enumerate}[label=(P1')]
\item 
 $D(Ax+b;F_{AX+b})=D(x;F_{X})$ for any random vector $X \in \mathbb{R}^d$, any $d\times d$ nonsingular matrix A and any $d-$vector $b$. \label{ass:depth1strong}
\end{enumerate}
This is the case, for example, in the definition of \cite{ZuoSerfling}. We refer to \cite{Mosler} for a discussion on this assumption. 
Applying depth functions to the voting framework, assumption \ref{ass:depth0} means that a permutation of candidates' indexes does not influence the final result. Assumption \ref{ass:depth1} indicates that grading on the scale $[-1, 1]$ should lead to the same result as grading on $[0, 20]$, for example. Assumption \ref{ass:depth1strong}, on the contrary, imposes stability when a plan transform is applied to the scatter plot. That is, if scatter plot 2 is obtained from scatter plot 1 applying a given plan transform, the deepest point 2 is obtained by applying the same plan transform to the deepest point~1. When considering grading matrices, it does not make sense to apply any plan transform. This justifies that we can use a weaker assumption to define depth functions.

In \ref{ass:depth2} various notions of symmetry are possible (namely, from the most constraining to the weakest, central symmetry, angular symmetry and halfspace symmetry); we refer to \cite{ZuoSerfling} for a discussion on this topic. This assumption ensures that the depth function is intuitive since it is maximal at the innermost point of the scatter plot, where the term innermost corresponds to a more or less constraining notion of symmetry. It goes together with assumption \ref{ass:depth3}, which imposes that as a point moves away from the innermost point, the depth function should decrease monotonically. 

Next, assumption \ref{ass:depth4} imposes that the depth function decreases to zero by convention. That is, the depth at a point infinitely far from the distribution must be equal to zero. 

Assumption \ref{ass:depth5} is rarely discussed in depth-functions literature. It claims that the deepest points must belong to the hyperplane containing the points. This assumption is very weak. To our knowledge, it is satisfied by all depth functions proposed in the literature. Note that \ref{ass:depth5} can be satisfied even if the points with maximal depth are not in the convex hull of the distribution support.

 Applications of depth techniques include, for example, robust estimation, center-outward ordering of multivariate observations, data exploration and multivariate confidence regions.
Several measures of data depth have been proposed in nonparametric statistics as  multidimensional generalizations of the ranks and median, each attempts to maintain certain robustness properties.

In this work, the deepest point is a location estimator of the preferences of the voters. The coordinates of the deepest point are the grades the innermost (possibly imaginary) voter would give.

A matrix of gradings $\Phi$ is a $n$-sample of a distribution $\Phi_n^*$ in $\R^d$, each profile $\Phi(\cdot,j)$, $j=1,\dots,n$ being a realization of $\Phi_n^*$. A sample version of depths can be defined using the empirical distribution.

For any grading matrix $\Phi$, denote by $\Phi_n$ the associated empirical distribution on $[0,1]^d$. That is, for $x=(x_1,\dots,x_d)\in\R^d$, $\Phi_n(x)=\left(\frac{1}{n}\sum_{j=1}^n \1\{\Phi(i,j)= x_i\}\right)_{i=1,\dots,d}$, 
where $\1\{A\}$ is equal to 1 if condition $A$ is satisfied and 0 else. The depth function is then applied on the empirical distribution $\Phi_n$. By abuse of notation, we will denote indifferently $D(\cdot;\Phi_n)$ or $D(\cdot;\Phi)$ the sample depths. Observe that, replacing $F$ by its empirical version, properties \ref{ass:depth2}, \ref{ass:depth3} and \ref{ass:depth4} may not be satisfied by the resulting sample depth. Nevertheless, the theoretical definition of the depth function ensures that it makes sense to consider the sample depth as an objective function. 

Let us give some examples of depth functions, which all satisfy Definition~\ref{def:depth}. The following descriptions deal with the empirical version of depth functions.
\begin{description}
\item[The weighted $L^p$ depths] \citep{Zuo2004} of a point $x\in \mathbb{R}^d$, $wL^pD(x;\Phi_n)$, given a set of $n$ points $\Phi(.,1),\hdots, \Phi(.,n)$ in $\mathbb{R}^d$ is defined by 
$$ w L^pD(x;\Phi_n)=\frac{1}{1+\frac{1}{n} \sum_{j=1}^n \omega( \|\Phi(.,j)-x\|_p)},  $$
where $p>0$, $\omega$ is a non-decreasing and continuous function on $[0,\infty)$ with $\omega(\infty)=\infty$
 and $\| x-x' \|_p=\left(\sum_{i=1}^d |x_i-x'_i|^p\right)^{1/p}$. As noted in \cite{Zuo2004}, the weighted $L^p$ depth is indeed a depth, as specified in Definition~\ref{def:depth}, for a distribution set $\mathcal F$ if for all $F\in\mathcal F$, for all $X\sim F$, $E w(\|x-X\|_p)<\infty$ for any $x\in\R^d$. In particular, this inequality holds when $w(x) =\sum_{k=0}^K a_k x^{b_k}$, $K\geq 0$, $a_K, b_K>0$, and $\forall k=0,\ldots, K-1$, $a_k, b_k\geq 0$. 
If $\omega: x \mapsto x^p$, then 
\begin{equation}
\label{eqn:wLp} L^pD(x;\Phi_n):= \frac{1}{1+\frac{1}{n} \sum_{j=1}^n   \sum_{i=1}^d |\Phi(i,j)-x_{i}|^p}
\end{equation}
will be called a $L^p$ depth.

For $p=\infty$, let us define also
\begin{equation}
\label{eqn:wLinf}
L^\infty D(x;\Phi_n):= \frac{1}{1+\frac{1}{n} \sum_{j=1}^n   \max_{i=1,\dots d} |\Phi(i,j)-x_{i}|}.
\end{equation}
Observe that $L^p$ depths do not satisfy assumption \ref{ass:depth1strong} \citep{Mosler}.

\item[The halfspace depth] \citep{Tukey} of a point $x\in \mathbb{R}^d$, $HD(x;\Phi_n)$, given a set of $n$ points $\Phi(.,1),\hdots, \Phi(.,n)$ in $\mathbb{R}^d$ is defined by 
\begin{align*} HD(x;\Phi_n):= & \textrm{ minimum proportion of voters}\\
&\textrm{ in a halfspace $H$ including } x.\end{align*}

\item[The projection depth] \citep{Zuo2003} of a point $x\in \mathbb{R}^d$, $PD(x; \Phi_n)$, given a set of $n$ points $\Phi(.,1),\hdots, \Phi(.,n)$ in $\mathbb{R}^d$, is defined by 
$$ PD(x; \Phi_n):= \inf_{u\in\R^d,\;\lVert u\rVert =1} \frac{1}{1+\lvert u^Tx -\mu(F_u)\rvert/\sigma(F_u)},$$
where $\lVert\cdot\rVert$ denotes the euclidean norm, exponent $T$ denotes the transpose operator, $\mu(F)$ denotes a central statistic of a distribution $F$ and $\sigma(F)$ a dispersion statistic. $F_u$ is the empirical distribution of $u^T \Phi$. In the following we will consider that $\mu(\cdot)$ is the median and $\sigma(\cdot)$ is the median absolute deviation.
Other choices are possible, such as the mean for $\mu(\cdot)$ and the standard deviation for $\sigma(\cdot)$. 
Here, the deepest point minimizes the maximal outlyingness in any given direction with respect to the scatter plot.

\item[Oja's depth] \citep{Oja} of a point $x\in \mathbb{R}^d$, $OD(x; \Phi_n)$, given a set of $n$ points $\Phi(.,1),\hdots, \Phi(.,n)$ in $\mathbb{R}^d$ is defined in \cite{Rousseeuw} by 
\[
OD(x; \Phi_n)=\frac{1}{1+ \sum_{(i_1, \ldots, i_d)} Volume~ S[x,\Phi(\cdot, i_1), \ldots , \Phi (\cdot, i_d)]}, 
\]
where $S[x_{i_1} , \ldots , x_{i_{d+1}}]$ is the closed simplex with vertices $x_{i_1}$, \ldots, $x_{i_{d+1}}$.
%\url{https://www.csun.edu/~ctoth/Handbook/chap58.pdf}, page 1545.}

\item[Weighted mean depths] \citep{WeightedMeans} are defined as follows. For all $\Phi\in\R^{d\times n}$, for all $\alpha\in(0;1]$, there exist positive weights $\{w_{j,\alpha},~j=1,\dots,n\}$, increasing in $j$ and with a sum equal to 1,  such that the depth $D$ satisfies \begin{multline*}
\{x\in\R^d, D(x; \Phi)\geq \alpha\} =\\ \mbox{conv}\left \{ \sum_{j=1}^n w_{j,\alpha}\Phi(\cdot,\pi(j)),~\text{with } \pi \text{ permutation of } \{1,\dots,n\}\right \}.
\end{multline*} 
This family includes include zono\"id depth, geometrical depth and expected convex hull\footnote{Geometrical depth and expected convex hull correspond respectively to weights equal to 
\begin{align*}
w_{j,\alpha}^\text{geometrical}&=\frac{1-\alpha}{1-\alpha^n}\alpha^{n-j}\1\{0<\alpha<1\}+\frac{1}{n}\1\{\alpha=1\},\\
w_{j,\alpha}^\text{ECH}&=\frac{j^{1/\alpha}-(j-1)^{1/\alpha}}{n^{1/\alpha}}.
\end{align*}
Zono\"id depth is obtained with weights $w_{j,\alpha}$ such that $0\leq w_{j,\alpha}\leq (n\alpha)^{-1}$ and $\sum_{j=1}^n w_{j,\alpha}=1$. For instance, one can consider \begin{equation}\label{eqn:zonoid}
w_{j,\alpha}^\text{zono\"id}=\frac{1}{n\alpha}\1\{j>n\alpha-\lfloor n\alpha \rfloor\}+\frac{1}{n\alpha}(n\alpha-\lfloor n\alpha \rfloor)\1\{j=n\alpha-\lfloor n\alpha \rfloor\}.
\end{equation}}.
The deepest point is obtained by letting $\alpha$ tend toward 1. As stated by \cite[Proposition 5]{WeightedMeans} the deepest point of a weighted mean depth is always the component-wise mean, that is, $\frac{1}{n}\sum_{j=1}^n\Phi(\cdot,j)$, which is also the deepest point for the $L^2$ depth.
\end{description}

Depths functions described above satisfy the assumptions \ref{ass:depth0}--\ref{ass:depth5}. 

We refer \emph{e.g.} to \cite{Mosler} or \cite{ZuoSerfling} for an overview on depth functions. Note that Liu's simplicial depth \citep{Liu}, even if well-known, does not satisfy Definition \ref{def:depth} since property \ref{ass:depth3} does not hold in general (see Counterexample 1 of \cite{ZuoSerfling}).
Let us now explain the link between a depth function and its associated social decision function.

\subsection{Deepest Voting}

In the following, the distribution of the grades $\Phi$ of the $n$ voters will be defined as $\Phi_n$, giving a 
weight $\frac{1}{n}$ at the independent profiles $\Phi(.,j)=(\Phi(1,j),\hdots,\Phi(d,j)) \in \mathbb{R}^d$,
$j = 1,\hdots,n$. Each profile can be seen as a point of $\R^d$ and, hence, depth functions can be applied to points $\Phi(.,j), j=1,\dots, n$.

\begin{figure}[!ht]
\centering
\includegraphics[height=8cm]{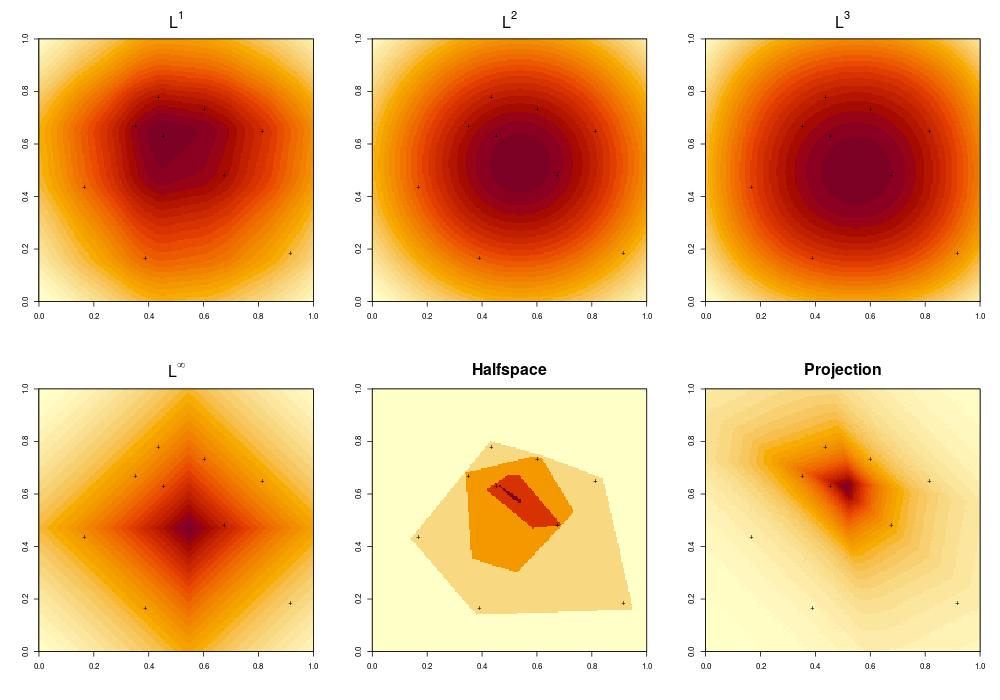}
\caption{Examples of depth functions relying on the example of Section \ref{sec:grades}. Horizontal axes give the grade for candidate $c_1$ and vertical axes for candidate $c_2$. Each cross corresponds to a voter.}
\label{fig:depths} 
\end{figure}

Figure \ref{fig:depths} illustrates the behavior of six usual depth functions on the example of Section \ref{sec:grades}. Namely it displays the values of  the $L^1$, $L^2$, $L^3$, and $L^\infty$ depths \citep{Zuo2004}, the halfspace depth \citep{Tukey}, and the projection depth \citep{Zuo2003}. 
 It highlights the diversity of depth measures.
 
\begin{definition}[Deepest Voting] 

Consider a grading matrix $\Phi$, and a given depth function $D$. Denote \[\mathcal X^*_D:= \{x\in\R^d : D(x,\Phi)=\sup(D(.,\Phi)) \}\]
the set of deepest points of $D$ with respect to $\Phi$. Let $G_D:\Phi\to \mathcal X^*_D$ be the method of grading with respect to the depth $D$.

Let \[i_D:=\argmax_{1 \leq i \leq d}  \{x^*_{D,i},~x^*_D=(x^*_{D,1},\hdots,x^*_{D,d})\in\mathcal X_D^\ast\}.
\]
The deepest voting process with respect to the depth $D$ is defined as the function which maps $\{\Phi(i,j),\; i=1,\dots,d,\; j=1,\dots,n\}$ to $i_D\subseteq \{1,\dots, d\}$.

If $i_D$ is unique, then the winner of the election is the candidate $c_{i_D}$. If $i_D$ is not unique, there is no unique winner of the election.
\end{definition}

As denoted by \cite{Zuo2013} the uniqueness of the deepest point may be acquired theoretically under symmetry assumptions on the distribution set $\mathcal F$ for many depths. Yet, in the sample case, the ideal symmetry situation seldom occurs, so uniqueness may often not be acquired. Some depths functions such as weighted $L^p$ depths with a strictly convex weight function or projection depths always admit a unique deepest point, that is  $\mathcal X^*_D=\{x^*_D\}$, but, for example, it is not necessarily the case for halfspace depth or Oja's depth \citep{Zuo2013}.  Table~\ref{tab:depths} recalls if  the uniqueness of the deepest point is satisfied or not for several depth functions.

It is worth noticing that even if $\mathcal X^*_D$ does not contain a unique element, the deepest voting $i_D$ may contain only one element. If $i_D$  contains several elements a tie-breaking rule should therefore be proposed. Such a rule can refer to the deepest space, {\it e.g.}, by reducing $\mathcal X^*_D$ to its unique center of gravity or refer directly to the candidates, {\it e.g.}, by electing the oldest candidate. Note that such tie-breaking rules are necessary for any voting process. See for instance \cite{Fabre} and references therein.

\begin{figure}[!ht]
\centering
\includegraphics[height=12cm]{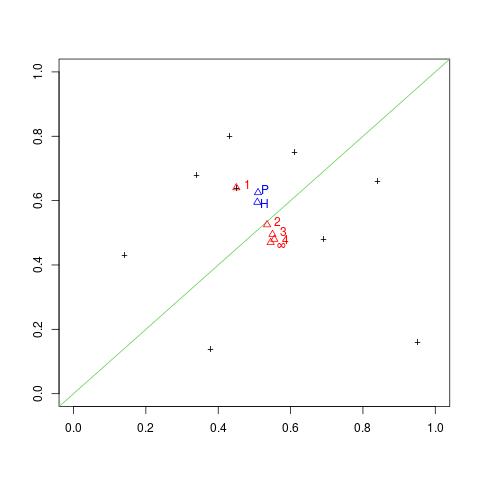}
\caption{ Examples of deepest points based on the example of Section \ref{sec:grades}. Horizontal axes give the grade for candidate $c_1$ and vertical axes for candidate $c_2$. Each cross corresponds to a voter. Triangles give the deepest points. Deepest points for the $L^p$ depths, for $p\in\{1,2,3,4,\infty\}$, are displayed with labels $p$. Deepest points for halfspace depth and projection depth are displayed with respective labels $H$ and $P$.}
\label{fig:nuage}
\end{figure}

Figure \ref{fig:nuage} displays some deepest voting results obtained on the grades given in the example of Section \ref{sec:example} for some depth functions. The transposed grading matrix $\Phi$ is represented as a scatter plot and deepest points are given in the figure. Note that the deepest point is not unique for halfspace depth; it was obtained by taking the center of gravity of the deepest set with respect to the Euclidean distance.

As seen in Figure~\ref{fig:nuage}, the deepest points do not have the same coordinates and they depend on the depth.
Table~\ref{tab:coord} gives the coordinates of the deepest points. Note that the deepest points of $L^1$, projection and halfspace depths designate the second candidate as the winner while the other deepest voting processes choose the first candidate.
 
\begin{table}[!ht]
\centering
\begin{tabular}{lcc} \hline
         & Candidate $c_1$  & Candidate $c_2$\\ \hline
$L^1$ depth  & 0.45 & 0.64\\
$L^2$ depth  & 0.54 & 0.52\\
$L^3$ depth  & 0.55 & 0.50\\
$L^4$ depth  & 0.56 & 0.48\\
$L^\infty$ depth  & 0.54 & 0.47\\ 
Halfspace depth & 0.51 & 0.60\\ 
Projection depth & 0.51 & 0.62\\ \hline
\end{tabular}
\caption{Coordinates of the deepest points of Figure~\ref{fig:nuage}.\vspace{\baselineskip} }
\label{tab:coord}
\end{table}

\section{Main properties of deepest voting}
\label{sec:conditions}

The objective of this section is to study if deepest voting processes satisfy \ref{C1}, \ref{C2}, \ref{C3}, \ref{C4}, and \ref{C5} as defined in Section~\ref{sec:conditionsgrading}.  

\subsection{Satisfaction of \ref{C1}, \ref{C2} and \ref{C3} properties}

First, \ref{C1}, \ref{C2}, \ref{C3} are satisfied by any depth function as defined in Definition~\ref{def:depth}, as stated by the following proposition.

\begin{proposition}  \label{prop:lien} 
Let $D$ be a depth function satisfying \ref{ass:depth0} and \ref{ass:depth5}. The deepest voting procedure $G_D$ associated to $D$ satisfies the properties \ref{C1}, \ref{C2}, and \ref{C3}.
\end{proposition}

\subsection{Satisfaction of \ref{C4} property}

\ref{C4} indicates that if a point of the scatter plot moves in a direction, the deepest point will not move in the opposite direction. We can establish that it is fulfilled by $L^p$ depth functions, and weighted mean depths family, which includes zono\"id depth, geometrical depth and expected convex hull depth.

\begin{proposition} \label{prop:condition4}
Let $D$ be either one of the following depths,
\begin{itemize}
\item a $L^p$ depth function with $1\leq p <\infty$, 
\item $L^\infty$ depth defined in \eqref{eqn:wLinf},
\item a depth in the weighted mean depths family.
\end{itemize}
Then the associated deepest voting procedure satisfies \ref{C4}. 
\end{proposition}

For the family of weighted mean depths, \ref{C4} follows from a property called \emph{monotonicity in the data} (\emph{e.g.} property T7 in \cite{WeightedMeans}). \ref{C4}, as formulated for a voting process, does not make sense when there is not a unique deepest point. In that case, a generalization can be expressed as follows. Suppose $\Phi = \widetilde\Phi$ except that one or more voters give higher grades to a candidate in $\Phi $ than in $ \widetilde\Phi $, then
\begin{equation} \label{eqn:c4set}
\argsup_{x\in\R^d} D(x;\Phi) \subseteq \{\argsup_{x\in\R^d} D(x;\widetilde\Phi)\}\oplus \R_+^d
\end{equation}
where $\oplus$ denotes the Minkowski sum of sets. The latter equation is satisfied by $L^1$ depth. 

Deepest voting procedures based on halfspace depth, projection depth, and Oja's depth seem less convenient since they do not satisfy \ref{C4}.

\begin{proposition} \label{prop:condition4halfspace}
The deepest voting procedures associated to halfspace depth, projection depth, and Oja's depth do not satisfy \ref{C4}. 
\end{proposition}

\subsection{Satisfaction of \ref{C5} property}

Next, there is no guarantee that a depth function $D$ satisfying assumptions \ref{ass:depth0}--\ref{ass:depth5} also fulfills \ref{C5}. For example, consider 7 voters and 3 candidates with the following grading matrices $\Phi$ and $\widetilde\Phi$:
\begin{align*} 
\Phi &=\begin{pmatrix}
  0.3 & 0.4 & 0.4 & 0.6 & 0.8 & 0.9 & 1\\
  0.7 & 0.6 & 0.6 & 0.4 & 0.2 & 0.1 & 0\\
  1   & 0.1   & 0.2   & 0.4   & 0   & 0   & 0
\end{pmatrix},\quad\\ \widetilde\Phi &=\begin{pmatrix}
  0.3 & 0.4 & 0.4 & 0.6 & 0.8 & 0.9 & 1\\
  0.7 & 0.6 & 0.6 & 0.4 & 0.2 & 0.1 & 0\\
  0   & 0   & 0   & 0.4 & 0.2 & 0.1 & 1
\end{pmatrix}.
\end{align*} \label{ex:iia}
Consider halfspace depth. It gives unique deepest points, which are respectively $x^*=(0.6, 0.4, 0.0)$
  for $\Phi$ and $\widetilde x^* =(0.43, 0.57, 0.05)$ for $\tilde\Phi$.
It shows that \ref{C5} is not satisfied, since the deepest points differ, giving as well different winners.

Depth functions with a component-wise definition lead to voting processes satisfying \ref{C5}. Let us introduce the following definition.

\begin{definition}%[Component-wise deepest set]
A depth function is said to have a component-wise deepest set if for all $\Phi\in\Lambda^{d\times n}$, for all $i\in\{1,\dots,d\}$, there exists $D_i$ such that for all $x^*\in\argmax_{x\in\R^d} D(x; \Phi)$, we have $x^*_{i}\in\argmax_{x\in\R} D_i(x; \Phi(i,\cdot))$. 
\end{definition}

We are now in a position to characterize the IIA behavior.

\begin{proposition}  \label{prop:IIA} 
Let $D:\mathbb{R}^d \times \mathcal{F} \mapsto \mathbb{R}$ be a depth function as in Definition~\ref{def:depth}. The Deepest Voting procedure $G_D$ satisfies \ref{C5} if and only if it has a component-wise deepest set. 
\end{proposition}
In particular, weighted mean depths and $L^p$ depths satisfy the conditions of Proposition~\ref{prop:IIA} and thus provide a voting procedure that meets \ref{C1}, \ref{C2}, \ref{C3}, \ref{C4}, and \ref{C5}.

As stated in Proposition \ref{prop:IIA}, \ref{C5} corresponds to decision functions that associate a score to each candidate independently to the grades obtained by others candidates. It thus imposes the subclass of depth functions with coordinates of deepest points obtained component-wise.  

\begin{rem}
We do not have equivalence between the IIA condition and the use of a component-wise depth function since \ref{C5} deals only with the deepest points. There exist counterexamples where the deepest point is obtained component-wise while the depth is not component-wise, such as for example the zono\"id depth. In this case, conditions of Proposition~\ref{prop:IIA} are satisfied, but the depth function is not component-wise.
\end{rem}

Satisfying the IIA hypothesis is an old widely discussed polemical issue when dealing with a voting process. One can refer to \cite{Osborne1976} and \cite{Benson2016} for past and recent arguments about the importance of satisfying or not the IIA. The main pro IIA argument is that a voting process satisfying IIA minimizes the risk of manipulations and strategic voting. The main cons arguments are that the choice of an election winner depends on the entire context, and that the presence (or absence) of a candidate in the ballot is an information that has to be taken into account. It is interesting to notice deepest voting can satisfy or not IIA depending on the chosen depth function.

We did not found any depth function in literature which leads to a deepest voting procedure which does not satisfy \ref{C5} and which satisfies \ref{C4}. Nevertheless, we can establish that the two properties are not equivalent.

\begin{proposition}\label{prop:C4sansC5}
Let $D$ be the depth function built as follows: 
\begin{align}\label{eqn:defZ}
D: &\mathbb{R}^d \times \mathcal{F} \to \mathbb{R}\\
   & (x, F)  ~~ \mapsto \begin{cases}
   D_Z(x, F) & \text{  if  }  D_Z(x; F) < 0.8\\
   1         & \text{  for } $x$ \text{ such that } \forall j=1,\dots, d\\
            & \qquad x_j=\mbox{mean}\{y_j,~ D_Z(y; F) \geq 0.8\}\\
   0.8       &\text{  elsewhere. }
   \end{cases}
\end{align}
where $D_Z$ denotes the zonoïd depth described in \eqref{eqn:zonoid}.
Then $D$ is a depth function satisfying \ref{ass:depth0} to \ref{ass:depth5}. The associated deepest voting procedure satisfies \ref{C1}, \ref{C2}, \ref{C3}, \ref{C4} but does not satisfy \ref{C5}.
\end{proposition}

\subsection{Synthesis}

Proposition~\ref{prop:IIA} highlights the link between depth function properties and the associated decision function. Considering a component-wise depth function ensures the IIA property. All the same, the choice of the depth function may be linked with the desired properties of the decision function. In particular, the more robust a depth function is, the less sensitive to extreme votes a decision function becomes. This is illustrated with the behavior of $L^p$ deepest points in Figure~\ref{fig:nuage}, where $L^1$ depth is more robust to extreme votes than $L^\infty$ depth.
Applying the propositions of this section to usual depth functions, we are able to characterize the behavior of several depth voting procedures. Table \ref{tab:depths} displays whether the unicity of the deepest point and the \ref{C5} are satisfied. Recall that \ref{C1}, \ref{C2} and \ref{C3} hold whatever the depth function. For usual depth functions in literature, either both \ref{C4} and \ref{C5} are satisfied, either none of them. Hence, Proposition \ref{prop:C4sansC5} shows that those properties are not equivalent and that it is possible to build deepest voting procedures satisfying \ref{C4} but not \ref{C5}. Others factors are to be considered, such as computational complexity (see \cite[Section 5]{Mosler2021} or \cite{aloupis2006}) and robustness (see \cite[Section 4.2]{Mosler2021} and references therein).

Note that the deepest points obtained by weighted mean depths and $L^2$ depth coincide. Hence, they are equivalent for the construction of a voting process. The use of one of the depth rather than another would only be useful if one wants to do another analysis, such as a profiling of the votes.
\begin{table}[!ht]
~\hspace{-1.5cm}{\small
\begin{tabular}{lccccccc}
Depth             & Unicity &\ref{C1} & \ref{C2} &\ref{C3} &\ref{C4}   & \ref{C5}\\ 
\hline 
$L^1$   & N  & $\checked$ & $\checked$ & $\checked$ & $\checked$ & $\checked$  \\
$L^p$ with $1< p \leq \infty$ & $\checked$  & $\checked$ & $\checked$ & $\checked$ & $\checked$ & $\checked$ \\
Halfspace depth      & N  & $\checked$ & $\checked$ & $\checked$ & N  & N \\ 
Oja depth            & N  & $\checked$ & $\checked$ & $\checked$ & N  & N \\
Projection depth     & $\checked$  & $\checked$ & $\checked$ & $\checked$ & N  & N \\
Weighted mean depths       & $\checked$  & $\checked$ & $\checked$ & $\checked$ & $\checked$  & $\checked$ \\
\end{tabular}
}
\caption{The table analyzes if the deepest voting procedure associated to several depth functions satisfies the uniqueness of the deepest vote and conditions defined in Section \ref{sec:conditionsgrading}, with $d>1$ candidates. $\checked$ corresponds to the verified properties, N to non verified ones. Recall that weighted mean depths and $L^2$ depth lead to the same voting process.}
\label{tab:depths}
\end{table}

Most of depth functions in literature were built to take into account the whole structure of the scatter plot and, hence, do not satisfy the IIA property. To obtain \ref{C5}, we can consider depth functions of the form $D(x;F)=\sum_{i=1}^d D^{(1)}(x_i,F_i)$ with $F_i$ marginal distribution of the multivariate distribution $F$, $D^{(1)}:\mathbb{R} \times \mathcal{F}^{(1)} \to \mathbb{R}$ a univariate depth function, and $\mathcal F^{(1)}$ the class of distributions on the Borel sets of $\mathbb{R}$. The sample version is $D(x;\Phi)=\sum_{i=1}^d D^{(1)}(x_i,\Phi(i,\cdot))$. A specific case is given by $D^{(1)}(x_i,\Phi(i,\cdot))=1/(1+\sum_{j=1}^n \rho(|x_i-\Phi(i,j)|)$ with $u\mapsto\rho(|u|)$ a non decreasing function. For instance, we can take $\rho$ equal to the truncated mean, Huber's, or Tukey's biweight loss functions. They are given respectively by 
\begin{gather*}
\rho_{\text{truncated}}(u)=\begin{cases} u^2 & \text{if } |u|\leq\lambda\\
\lambda^2 &\text{else},\end{cases} \qquad
\rho_{\text{Huber}}(u)=\begin{cases} u^2/2 & \text{if } |u|\leq\lambda\\
\lambda u -\lambda^2/2 &\text{else}
\end{cases}, \\
{\rho_{\text{Tukey}}(u)=\begin{cases} \lambda^2(1-(1-(u/\lambda)^2)^3)/6 & \text{if } |u|\leq\lambda\\
\lambda^2/6 &\text{else},
\end{cases}}
\end{gather*}
with $\lambda>0$. 
For $\lambda$ sufficiently large, these latter choices leads to depth functions with a unique deepest point, and satisfying \ref{C1}, \ref{C2}, \ref{C3}, \ref{C4}, and \ref{C5}. This is a generalization of $L^p$ depths.

\begin{rem}
A paradox which may appear in vote processes is the election of a candidate with stochastically dominated grades\footnote{A candidate $c_1$ dominates stochastically a candidate $c_2$ if for all $v\in(0,1)$, there are more grades higher than $v$ for candidate $c_1$ than for candidate $c_2$.}. \cite{zuo2000performance} observes that this paradox is linked with the symmetry property in condition \ref{ass:depth2}. For example, it occurs with halfspace depth, projection depth and Oja's depth, but not with range voting and majority judgment.
\end{rem}

\begin{rem} Seeing a voting procedure as an optimization problem, deepest voting can also be interesting in some contexts. Consider, for example, a vote on a budget plan. The candidates are, here, sectors in which the budget must be distributed. Voters associate with the sectors the portion of the budget that they wish to allocate, so that the sum of the allocations is equal to 100\%. Deepest voting approach makes it possible to find the budget by maximizing an objective function under the constraint that the sum of the portions must be equal to 100\%.
\end{rem}

\begin{rem}
Component-wise depth functions may be able to consider partial abstention, evaluating each coordinate of the deepest point by only considering non-missing grades. Yet, a non component-wise depth function will have to remove all voters with partial abstention from the procedure. All the same, it does not make sense to consider a vote procedure which is not IIA on partial votes. Another possibility is to do imputation of missing values \citep{imputation}. Yet, from a political point of view, not considering missing grades may be more appropriate.
\end{rem}

\begin{rem} 
Taking into account grading information in a voting process leads to the notion of sensibility to extreme evaluations. With deepest voting procedures, this sensibility can be measured by the breakdown point of the associated depth function. As defined by \cite{Zuo2004}, ``Roughly speaking, the finite sample breakdown point of an estimator is the minimum fraction of `bad' points in a data set that can render the estimator useless. In the location setting, if the estimator becomes unbounded under some contamination, then we say the estimator becomes useless''. The maximal value of a breakdown point is 1/2.
In the voting context, a deepest point cannot be unbounded, but the breakdown point highlights the robustness of a procedure to extreme values. 
We recall below the breakdown points of deepest points for some classical depths.

\begin{tabular}{ll}
depth & breakdown point\\ \hline
 $L^1$ depth & $1/2$\\ 
 $L^p$ depth with $p>1$ & $1/n$\\ 
 Halfspace depth & $1/(d+1)$\\ 
 Projection depth & $1/2$\\ 
 Oja's depth & $2/n$\\ 
Weighted mean depths & $1/n$
\end{tabular}

For more details, we refer the reader to \cite{Rousseeuw}, \cite{Zuo2004} and \cite{ZuoSerfling}. The higher the breakdown point is, the less the procedure is sensitive to extreme grades, which means that the procedure is less manipulable.
\end{rem}

\section[Properties of Lp deepest voting family]{Properties of $L^p$ deepest voting family}
 %----------------------------------------------------------------------------------------
\label{sec:wLp}

In the following, we focus on the family of $L^p$ depths. This choice is stimulated by the fact that usual social decision functions are members of this family. We assume that we dispose of a set of $n$ points $\Phi(.,1),\hdots, \Phi(.,n)$ in $\mathbb{R}^d$. As described above, the $L^p$ depth \citep{Zuo2004} of a point $x$, $L^p D(x,\Phi)$, is defined by \eqref{eqn:wLp} and \eqref{eqn:wLinf}.
The {$L^p$ deepest voting} takes the grading matrix $\Phi$  as argument and returns the coordinates of the point maximizing the $L^p$ depth function applied to $\Phi$.

We can first check that $L^p$ deepest voting defines, indeed, a decision function in the sense of \cite{Balinski2007}.
\begin{proposition} \label{p1}
For $p >1$, let us consider the associated $L^p$ deepest voting and denote by $D$ the depth function. Then the \emph{method of grading} $G_D$ satisfies \ref{C1}, \ref{C2}, \ref{C3}, \ref{C4}, and \ref{C5}, which are exposed in Subsection~\ref{sec:conditionsgrading}.

For $p=1$, the \emph{method of grading} $G_D$, associated to $L^1$ deepest voting, satisfies \ref{C1}, \ref{C2}, \ref{C3}, and \ref{C5} exposed in Subsection~\ref{sec:conditionsgrading}. The generalization \eqref{eqn:c4set} of \ref{C4} is also verified. 
\end{proposition}
This proposition is a corollary of Proposition~\ref{prop:lien}, Proposition~\ref{prop:condition4} and Proposition~\ref{prop:IIA}.

The next propositions deal with the characterization of the weighted $L^p$ deepest voting for different choices of $p$.
\begin{proposition}  \label{p1bis} 
For all $p>1$, the set of $L^p$ deepest points has a unique element.
For all $ 0 < p \leq 1$, the cardinal of the set of $L^p$ deepest points may be greater than 2.
\end{proposition}
The case $p<1$ should be avoided, as the uniqueness of the deepest point is not ensured. It can also be shown that if $p$ tends to 0, each voter is a deepest point!

In the case $p=1$ : 
\begin{itemize}
\item if $n$ is odd, then the deepest point is unique.
\item if $n$ is even, the set of the deepest points is composed by the points $x^*=(x^*_1,\hdots,x^*_d)$ such that for all $i=1,\dots,d$,  
\[ x^*_i \in [\Phi(i,\cdot)_{(n/2)};\Phi(i,\cdot)_{(n/2+1)}] \]
where the $\Phi(i,\cdot)_{(k)}$ is the $k^{th}$ observation of the set of the ordered voters' preferences for candidate $c_i$.
\end{itemize}

The class of $L^p$ deepest voting includes three usual voting processes, which are majority judgment \citep{Balinski2007}, approval voting \citep{Brams} and range voting \citep{Smith}.
\begin{proposition} \label{p2}
Without considering the tie-breaking procedure, the majority judgment belongs to the $L^1$ deepest voting set. Analogously, the range voting and the approval voting are obtained by the $L^2$ deepest voting.
\end{proposition}

$L^1$ deepest voting leads to the majority judgment. This method has many advantages (see  \cite{Balinski2007} for example). Nevertheless, it presents also several drawbacks (see \cite{Laslier2019} or \cite{Felsenthal2008}). A peculiar property of the $L^1$ deepest voting is that the value of the decision function may not be included in the convex hull of the $n$ voters' grades, \emph{i.e.}, $\mathcal X^*_D$ is not included in the convex hull of $\Phi$. Consider \emph{e.g.} the situation with 3 candidates and 3 voters and the associated matrix $\Phi$:
\begin{eqnarray*}
\Phi(1,.)=& (a,0,0) \\
\Phi(2,.)=& (0,b,0)\\
\Phi(3,.)=& (0,0,c)
\end{eqnarray*}
with $0 < a < b < c \leq 1$. The $wL^1$ deepest point is $x^*=(0,0,0)$, which is not included in the convex hull of the points $(a,0,0)$, $(0,b,0)$ and $(0,0,c)$.

Another case where we can explicit the deepest point is the $L^\infty$ deepest voting.
\begin{proposition}  \label{p3} 
$L^\infty$ deepest voting maps to the point
 whose coordinates are in the middle of the most extreme coordinates component-wise, that is, it
attributes to each candidate the mean of their best and worst grades.
\end{proposition}

\begin{rem}For $p>1$, the unicity of the deepest point implies that the probability of having tied-winners is low. When $p=1$, based on $L^p$ depths, a natural tie-breaking rule for majority judgment can be proposed. Suppose that two candidates have the same result for the majority judgment (obtained with a $L^1$ deepest voting), then one could elect the winner (if it exists) of the $L^p$ deepest voting when $p$ is strictly greater than 1 but tends to 1. 

This rule is different from the tie-breaking rule proposed by \cite{Balinski2020}. 
For a counterexample, consider five voters and two candidates $c_1$ and $c_2$ with respective grades in [0;1] $(0.45,0.45,0.5,1,1)$ and $(0.5,0.5,0.5,0.5,0.5)$.

Majority judgment (see \eqref{eqn:mj}) leads to the same final grade 0.5 for both of them. According to the Balinski-Laraki's tie-breaking procedure, one should remove a grade equal to the final result (here, 0.5) for both candidates, and apply majority judgment on the remaining grades. In this setting, it consists in applying majority judgment to $(0.45,0.45,1,1)$ and $(0.5,0.5,0.5,0.5)$. With this procedure, candidate $c_2$ should be elected since lowest-middlemost grade of $c_1$ is equal to 0.45.
Now, it can be shown that the derivative of $L^p$ relative to the first candidate at point 0.5 has the same sign as $0.05^{p-1}-0.5^{p-1}$. For every $p>1$, this quantity is clearly negative so, by convexity, the point maximizing the $L^p$ depth for $c_1$ is greater than 0.5. Since candidate $c_2$ has a constant grade 0.5, it is straightforward that the coordinate of the deepest point is also 0.5. Hence candidate $c_1$ is elected here. So, the tie-breaking rule of Balinski-Laraki method can't be seen as a natural limit of the criterion when p$\rightarrow$1.
\end{rem}

The choice of $p$ in $L^p$ deepest voting is of course critical. Roughly speaking, the greater $p$ is, the more importance is given to immoderate grades. 
For a large enough $p$, the deepest point component-wise depends only on most immoderate grades.
 So, the $L^p$ Deepest Voting is very vulnerable to strategic voting for a large $p$. These considerations meet the work of \cite{Balinski2007} arguing that majority judgment ($L^1$ Deepest Voting) is more resistant to strategic voting than range voting ($L^2$ Deepest Voting).

As pointed out above, any voting process suffers from unwanted properties generally named paradoxes (see \cite{Felsenthal}  for a more detailed description of voting paradoxes). As other voting processes,  {\em $L^p$ deepest voting} seems to be affected by paradoxes. We propose, in the following, to focus specifically on four classical properties and show how the paradigm changes due to the use of a grading model. The four studied properties are:
\begin{itemize}
\item  the \emph{Condorcet winner paradox}: a candidate $c$ is not elected despite the fact that $c$ is preferred by the majority of the voters over each of the remaining candidates;
\item  the \emph{Condorcet loser paradox}: a candidate $c$ is elected despite the fact that the majority of voters prefer each of the remaining candidates to $c$;
\item  the \emph{reinforcement paradox}: if $c$ is elected in each of several disjoint electorates, it is possible that $c$ is not elected if all electorates are combined into a single electorate;
\item  the \emph{no-show paradox}: a voter may obtain a more preferable outcome if he decides not to participate in an election than if he decides to participate in the election and vote sincerely for their preferences.
\end{itemize}

\begin{proposition}  \label{p4} 
For all $p \geq 1$, the $L^p$ deepest voting suffers from the Condorcet winner and the Condorcet loser paradoxes.
\end{proposition}

\begin{proposition}  \label{p5} 
$L^p$ deepest voting suffers from reinforcement and no-show paradoxes, for $p\in[1,\infty]\setminus{\{2\}}$. These properties do not hold for $L^2$ deepest voting.
\end{proposition}

Many properties which appear as paradoxes in ranking-based voting processes, with binary $\Lambda$, may not be seen as drawbacks under a more complex grading-based model, that is, for discrete or continuous sets $\Lambda$. The four properties studied above justify our statement. 

Consider for example the Condorcet winner paradox. Suppose that 3 voters $v_1, v_2, v_3$ have to choose between two candidates $c_1$ and $c_2$ with the following grades: 
\begin{center}
\begin{tabular}{lcc} \hline
         & Candidate $c_1$  & Candidate $c_2$\\ \hline
voter $v_1$  & 0.8 & 0.7\\
voter $v_2$  & 0.8& 0.7\\
voter $v_3$  & 0.1 & 0.9\\ \hline
\end{tabular}
\end{center}
In this configuration, the Condorcet winner is $c_1$. Yet, as we have a quantification of the preference of voters, we can see that candidate $c_2$ is less divisive in the sense that no voters reject this candidate with very low grades. The fact that a voting process elects $c_2$ is acceptable in this configuration.

Now consider the no-show paradox. Suppose that two candidates $c_1$ and $c_2$ obtained  the following grades for 3 voters:
\begin{center}
\begin{tabular}{lcc} \hline
         & Candidate $c_1$  & Candidate $c_2$\\ \hline
voter $v_1$  & 0.5 & 1\\
voter $v_2$  & 0.5& $\varepsilon$\\
voter $v_3$  & 0 & $\varepsilon$\\ \hline
\end{tabular}
\end{center}
Then, candidate $c_2$ is elected with voters $\{v_1, v_2\}$. But for some $\varepsilon>0$ sufficiently small, the candidate $c_2$ is not elected with voters $\{v_1, v_2, v_3\}$, considering $L^p$-deepest voting with $p<2$ (see the proof of Proposition~\ref{p5}). In this configuration, voter $v_3$'s votes will make $c_2$ lose. $L^p$ depth functions with $p<2$ favor candidates with less dispersion, which are more consensual. The no-show paradox results from the fact that $v_3$ does not approve candidate $c_2$ while their preference with respect to candidate $c_1$ is not significant.

When $p>2$, a configuration where no-show paradox holds is the following:
\begin{center}
\begin{tabular}{lcc} \hline
         & Candidate $c_1$  & Candidate $c_2$\\ \hline
voter $v_1$  & 0 & 0.5+$\varepsilon$\\
voter $v_2$  & 1 & 0.5+$\varepsilon$\\
voter $v_3$  & 0 & $\varepsilon$\\ \hline
\end{tabular}
\end{center}
with $\varepsilon>0$ sufficiently small (see the proof of Proposition~\ref{p5}). It is due to the low grade given by $v_3$ to $c_2$ and the fact that the difference with the grade for $c_1$ is not significant. $L^p$ depth functions with $p>2$ here favor candidates with highest grades. In both situations, the fact that no-show paradox occurs is acceptable since even if voter $v_3$ prefers candidate $c_2$, the grades show that the latter has not gained yet the support of voter $v_3$.

These examples illustrate that the amount of information available through grading is likely to change the result of a voting process. It also helps to explain some results which may be seen as paradoxes with preferential grades.

\section{Conclusion}

Grading-based voting offers a richer information than ranking-based model. It relies on discrete or continuous evaluations of candidates by voters rather than a single preference. Grading widens the scope of decision processes available. 

In this paper, we introduce a new viewpoint on voting procedures, by integrating a parallelism with the statistical notion of depth. Depth functions give a unified way to define voting processes, with finite or continuous grading. We define the notion of deepest voting including classical voting procedures such as majority judgment, approval voting and range voting.

We study the relation between the properties of the depth functions and that of the voting processes. We show that main usual properties of voting processes are satisfied, namely non-dictatorship, universality and unanimity for deepest voting associated to all classical depths. The monotonicity of the voting processes is proved for several  functions, such as  $L^p$ depths. Yet, we show that voting processes based on some usual depth functions, such as halfspace, projection or Oja's depths, do not satisfy monotonicity. The non-satisfaction of this property seems a main drawback for these deepest voting processes. We finally establish that independence to irrelevant alternatives is related with a component-wise behavior of the deepest point. 

Deepest voting related to the family of $L^p$ depths satisfies Balinski and Laraki's axioms. Range voting and approval voting are associated to $L^{2}$ depth and Majority judgment to $L^{1}$ depth.
We studied some basic paradoxes, namely Condorcet's, no-show and reinforcement paradoxes. We show that these paradoxes occur with this family of decision processes.
We provide a discussion on these paradoxes, where we highlight the changes generated by the use of a grading model. 
 
Deepest voting has paved the way for new voting procedures, by varying the depth function used in the voting process. Even if some classical depth functions seem inappropriate due to lack of monotonicity for example, deepest voting deserves much attention.  Depth functions have encountered much interest in statistical literature and their attractive properties (in particular robustness) may be useful in social decision theory.

\subsection*{Acknowledgments}

The authors would like to thank the anonymous referee and the associate editor for their comments that led to substantial improvements in the article. 
The authors also thank Rainer Dyckerhoff for valuable private communications about monotonicity.

\appendix

\section{Proofs}

%%%%%%%%%%%%%%%%%%%%%%%%%%%%%%%%%%%%%%%%%%%%%%%%%%%%%%%%%%%%%%%%%%%%%%%%%%%%%%%%%%%%%%%%%%

\subsection{Proof of Proposition \ref{prop:lien}}

\ref{C1} directly follows from assumption \ref{ass:depth0}.

Permuting the columns of $\Phi$ does not change the empirical distribution $\Phi_n$. Thus \ref{C2} is straightforward.

\ref{C3} directly follows from assumption \ref{ass:depth5}.

%%%%%%%%%%%%%%%%%%%%%%%%%%%%%%%%%%%%%%%%%%%%%%%%%%%%%%%%%%%%%%%%%%%%%%%%%%%%%%%%%%%%%%%%%%

\subsection{Proof of Proposition \ref{prop:condition4}}

Let $\Phi$ and $\widetilde\Phi$ be two grading matrices in $R^{d\times n}$. Suppose that $\Phi = \widetilde\Phi$ except that voter 1 gives an higher grade to candidate $c_1$ in $ \Phi $ than in $\widetilde \Phi$, that is,
\[
\forall i=1,\ldots, d, ~\forall j=1,\ldots , n, ~ (i,j)\neq(1,1), ~\Phi(i,j)= \widetilde\Phi (i,j) 
\]
and $\Phi(1,1)\geq \widetilde\Phi (1,1)$. We want to prove that if there exist unique deepest points $\theta$ and $\widetilde\theta$, associated respectively to $\Phi$ and $\widetilde\Phi$, then $\theta_1\geq \widetilde\theta_1$.

Without loss of generality we consider the case where only one vote is different between $\Phi$ and $\widetilde\Phi$. The reasoning can be done iteratively if more than one vote change for a candidate.

\subsubsection*{$L^p$ depths with $1\leq p<\infty$}

Consider the $L^p$ depths, $1< p<\infty$, defined for $x\in \mathbb{R}^d$ by 
\[L^pD(x;\Phi)=\frac{1}{1+\frac{1}{n}\sum_{j=1}^n \|x-\Phi(\cdot,j)\|_p^p}.\]
Let $\theta\in\argmax_{x\in\R^d} L^pD(x;\Phi)$ and $\widetilde\theta\in\argmax_{x\in\R^d} L^pD(x;\widetilde\Phi).$ 

It is easily seen that 
\begin{align*}
\theta &= \argmin_{x\in\R^d} \sum_{i=1}^d\sum_{j=1}^n \lvert\Phi(i,j)-x_i\rvert^p, \\
\widetilde \theta &= \argmin_{x\in\R^d} \sum_{j=1}^n \lvert \widetilde\Phi(1,j)-x_1\rvert^p + \sum_{i=2}^d\sum_{j=1}^n \lvert\Phi(i,j)-x_i\rvert^p.
\end{align*}
It follows that
$\theta_1 \in \argmin_{x_1\in\R} L(x_1,\Phi)$, and $\widetilde \theta_1 \in \argmin_{x_1\in\R} L(x_1,\widetilde\Phi)$, with
$L(x_1,\Phi)=\sum_{j=1}^n |\Phi(1,j)-x_1|^p $ and
\[L(x_1,\widetilde\Phi)= \sum_{j=1}^n |\widetilde\Phi(1,j)-x_1|^p =L(x_1,\Phi)+|\widetilde\Phi(1,1)-x_1|^p-|\Phi(1,1)-x_1|^p.\]
The functions $x\mapsto L(\cdot;\Phi)$ and $x\mapsto L(\cdot;\widetilde\Phi)$ are convex and almost surely differentiable.

We distinguish with respect to the values of $p$.
\begin{itemize}
\item If $p>1$,\\
The functions $L(\cdot;\Phi)$ and $L(\cdot;\widetilde\Phi)$ are strictly convex.  Hence, $\theta_1$ and $\widetilde\theta_1$ are uniquely defined. It suffices to show that $L'(\theta_1;\widetilde\Phi)\geq 0$ to deduce $\theta_1 \geq \widetilde\theta_1$.
\item If $p=1$,\\
The functions $L(\cdot;\Phi)$ and $L(\cdot;\widetilde\Phi)$ are convex, but not strictly convex, and the points attaining the minimum are possibly not unique (when $n$ is even).
Showing that  $L'(\theta_1;\widetilde\Phi)\geq 0$ implies that either $\theta_1\in\argmin_{x_1\in\R} L(x_1;\widetilde\Phi)$ or $\theta_1\geq \widetilde x_1^*$ for all $\widetilde  x_1^*\in \argmin_{x_1\in\R} L(x_1;\widetilde\Phi)$. In that case, we obtain the generalization \eqref{eqn:c4set} of \ref{C4}.
\end{itemize}
Consequently, it suffices for our purposes to prove that $L'(\theta_1;\widetilde\Phi)\geq 0$.

First observe that since $\theta_1$ minimizes $L(\cdot;\Phi)$, $L'(\theta_1,\Phi)=0$. Hence, we have
\begin{multline*}
L'(\theta_1,\widetilde\Phi)= p|x_1-\widetilde\Phi(1,1)|^{p-1}\sign(\theta_1-\widetilde\Phi(1,1))\\
- p|\theta_1-\Phi(1,1)|^{p-1}\sign(\theta_1-\Phi(1,1)).
\end{multline*}
We distinguish three cases.
\begin{itemize}
\item If $\theta_1-\widetilde\Phi(1,1)\leq0$,~\\
then, $\theta_1-\Phi(1,1)\leq\theta_1-\widetilde\Phi(1,1)\leq 0$. It results that  $L'(\theta_1,\widetilde\Phi)\geq 0$.
 
\item If $\theta_1-\widetilde\Phi(1,1)>0$ and $\theta_1-\Phi(1,1)<0$,~\\
then, $L'(\theta_1,\widetilde\Phi)\geq 0$.
 
\item If   $\theta_1-\Phi(1,1)\geq 0$,~\\
then, $\theta_1-\widetilde\Phi(1,1)\geq \theta_1-\Phi(1,1)\geq 0$. It results that $L'(\theta_1,\widetilde\Phi)\geq 0$.
\end{itemize}
This concludes the proof.

\subsection*{$L^\infty$ depth} 

Consider the $L^\infty$ depth, which is defined for $x\in \mathbb{R}^d$ by 
\[L^\infty D(x;\Phi)=\frac{1}{1+\frac{1}{n}\sum_{j=1}^n \|\Phi(\cdot,j)-x\|_\infty }.\]
Let $\theta=\argmax_{x\in\R^d} L^\infty D(x;\Phi)$ and $\widetilde\theta=\argmax_{x\in\R^d} L^\infty D(x;\widetilde\Phi).$
Then, Proposition~\ref{p3} establishes that, for all $i=1,\dots,d$,
\begin{align*}
\theta_i &= \frac{\min_{j=1,\dots,n} \Phi(i,j)+\max_{j=1,\dots,n} \Phi(i,j)}{2},\\
\widetilde\theta_i &= \frac{\min_{j=1,\dots,n} \widetilde\Phi(i,j)+\max_{j=1,\dots,n} \widetilde\Phi(i,j)}{2}.\\
\end{align*}
Hence, for all $i=2,\dots,d$, $\theta_i=\widetilde\theta_i$. 
Since $\widetilde \Phi(1,1)\leq \Phi(1,1)$, we have $\max_{j=1,\dots,n} \widetilde \Phi(1,j)\leq \max_{j=1,\dots,n} \Phi(1,j)$. We next distinguish two cases.
\begin{itemize}
\item If $\min_{j=1,\dots,n} \Phi(1,j)\leq \widetilde\Phi(1,1)$, we get $\min_{j=1,\dots,n} \Phi(1,j)=\min_{j=1,\dots,n} \widetilde\Phi(1,j)$ and, hence, $\widetilde\theta_1 \leq \theta_1$.
\item If $\min_{j=1,\dots,n} \Phi(1,j)> \widetilde\Phi(1,1)$, we get $\widetilde\theta_1=\frac{\widetilde\Phi(1,1)+\max_{j=1,\dots,n} \widetilde\Phi(1,j)}{2}\leq \theta_1$.
\end{itemize}
The proof is complete.

\subsection*{Weighted mean depths}

Recall a result stated in \cite[page 26]{Mosler} for weighted mean depths family.
Let $D$ be a depth function in the weighted mean depths family.
Denote $D_\alpha(\Phi)=\{z \in \R^d,~ D(z;\Phi) \geq \alpha \}$.  
If $\Phi(i,j)\geq \widetilde\Phi(i,j)$ holds for all $i=1,\dots,d$, $j=1,\dots,n$ then \[
D_\alpha(\Phi) \subseteq D_\alpha (\widetilde\Phi ) \oplus \R^d_+,
\]
where $\oplus$ denotes the Minskowski sum of sets. Moreover we have unicity of the deepest point.
Consequently, \ref{C4} is satisfied.

%%%%%%%%%%%%%%%%%%%%%%%%%%%%%%%%%%%%%%%%%%%%%%%%%%%%%%%%%%%%%%%%%%%%%%%%%%%%%%%%%%%%%%%%%%

\subsection{Proof of Proposition \ref{prop:condition4halfspace}}

It is sufficient to exhibit two grading matrices $\Phi$ and $\widetilde\Phi$ with equal grades except that $\widetilde \Phi$ has a higher entry for candidaiate $c_1$, such that the first coordinate of the deepest point for $\Phi$ is lower than the first coordinate of the deepest point for $\widetilde\Phi$. Let us distinguish with respect to the depth functions.

\subsubsection{Halfspace depth}

Consider 8 voters and 2 candidates with the following grading matrices $\Phi$ and $\widetilde\Phi$:
\begin{align}\label{eqn:half} 
\Phi&=\begin{pmatrix}
  0 & 0.0 & 0.0 &  0.2 & 0.6 & 0.8 & 1.0 & 0.3 \\
  0 & 0.2 & 0.4 & 1.0 & 0.0 & 1.0   & 0.6 & 0.4 
\end{pmatrix},\\
\widetilde\Phi&=\begin{pmatrix}
   0 & 0.0 & 0.0 &  0.2 & 0.6 & 0.8 & 1.0 & 0.4 \\
  0 & 0.2 & 0.4 & 1.0 & 0.0 & 1.0   & 0.6 & 0.4 
\end{pmatrix}.
\end{align}
The corresponding deepest points obtained by the halfspace depth are equal respectively to $x^*=( 0.3, 0.4)$ and $\widetilde x^*=(0.28, 0.38)$.
This is illustrated in Figure \ref{fig:half} below.

\begin{figure}[!ht]
\centering
\includegraphics[height=6cm]{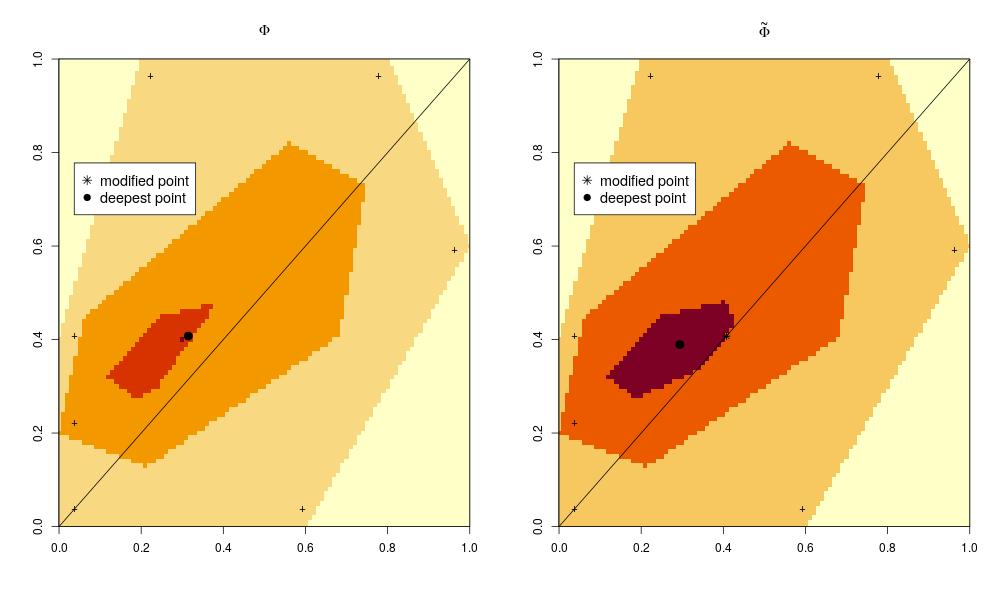}
\caption{ Deepest points based on the halfspace depth for grading matrices defined in \eqref{eqn:half}. Horizontal axes give the grade for candidate $c_1$ and vertical axes for candidate $c_2$. Each cross corresponds to a voter.}
\label{fig:half}
\end{figure}

\subsubsection{Projection depth}

Consider 8 voters and 2 candidates with the following grading matrices $\Phi$ and $\widetilde\Phi$:
\begin{align}\label{eqn:proj} 
\Phi&=\begin{pmatrix}
  0.0 & 0.1 & 0.4 & 0.5 & 0.5 & 0.5 & 0.5 & 0.5 \\
  0.1 & 0.2 & 0.5 & 0.6 & 0.1 & 0.2 & 0.4 & 0.8
\end{pmatrix},\\ 
\nonumber\widetilde\Phi&=\begin{pmatrix}
 0.0 & 0.1 & 0.4 & 0.5 & 0.5 & 0.5 & 0.5 & 0.7\\
 0.1 & 0.2 & 0.5 & 0.6 & 0.1 & 0.2 & 0.4 & 0.8
\end{pmatrix}.
\end{align}
The corresponding deepest points obtained by the projection depth are equal respectively to $x^*=(0.5, 0.5)$ and $\widetilde x^*=(0.44, 0.54)$.
This is illustrated in Figure \ref{fig:proj} below.

\begin{figure}[!ht]
\centering
\includegraphics[height=6cm]{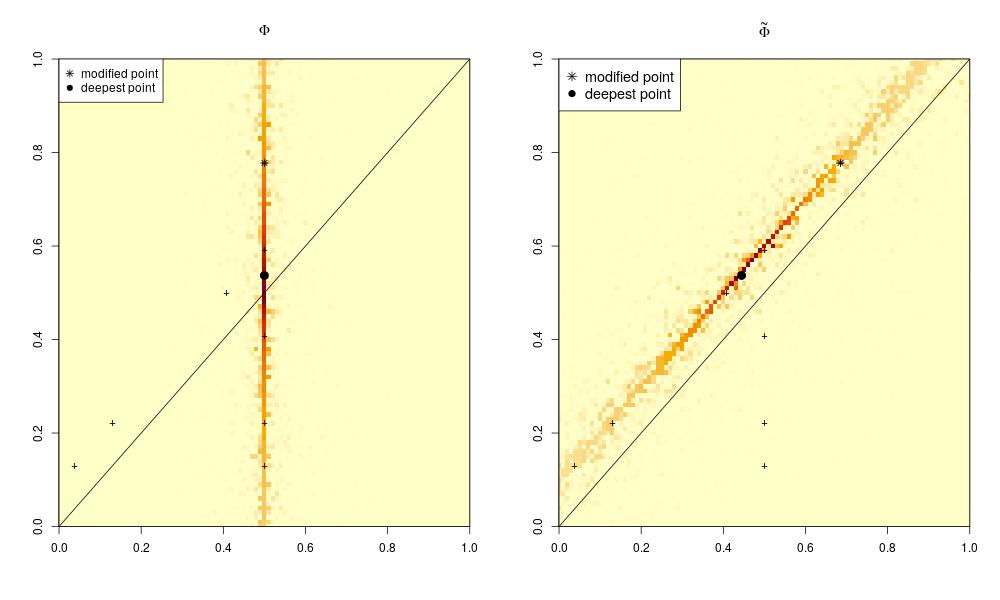}
\caption{ Deepest points based on the projection depth for grading matrices defined in \eqref{eqn:proj}. Horizontal axes give the grade for candidate $c_1$ and vertical axes for candidate $c_2$. Each cross corresponds to a voter.}
\label{fig:proj}
\end{figure}

\subsubsection{Oja's depth}

Consider 8 voters and 2 candidates with grading matrices $\Phi$ and $\widetilde\Phi$ given by \eqref{eqn:proj}.
The corresponding deepest points obtained by Oja's depth are equal respectively to $x^*=(0.5, 0.4)$ and $\widetilde x^*=(0.5, 0.5)$.
This is illustrated in Figure \ref{fig:oja} below.

\begin{figure}[!ht]
\centering
\includegraphics[height=6cm]{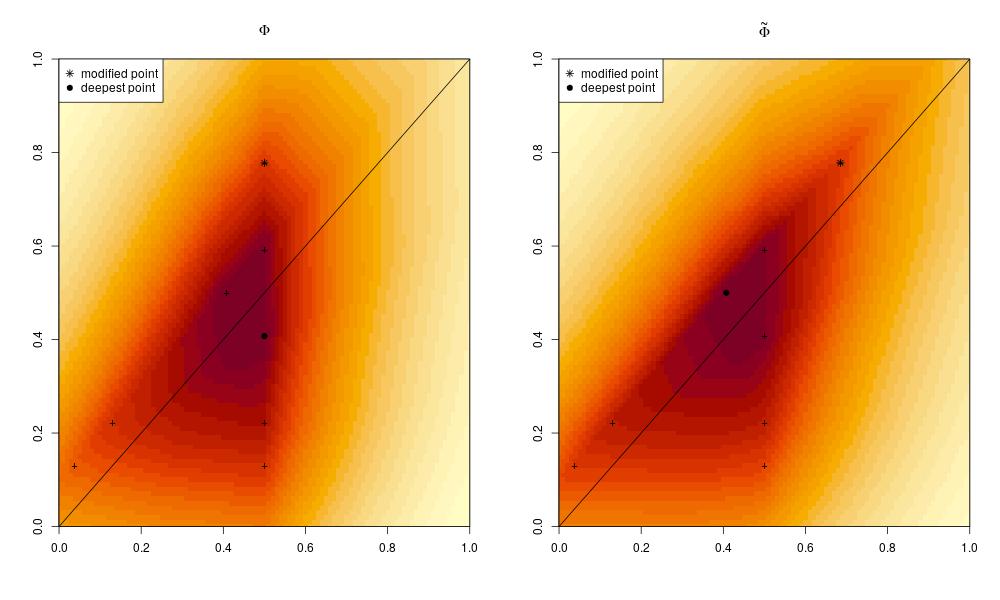}
\caption{ Deepest points based on Oja's depth for grading matrices defined in \eqref{eqn:proj}. Horizontal axes give the grade for candidate $c_1$ and vertical axes for candidate $c_2$. Each cross corresponds to a voter.}
\label{fig:oja}
\end{figure}

%%%%%%%%%%%%%%%%%%%%%%%%%%%%%%%%%%%%%%%%%%%%%%%%%%%%%%%%%%%%%%%%%%%%%%%%%%%%%%%%%%%%%%%%%%

\subsection{Proof of Proposition \ref{prop:IIA}}

Suppose first that there exists $D_1$ such that for all $x^*\in\argmax_{x\in\R^d} D(x; F)$,  $x^*_{1}\in\argmax_{x\in\R} D_1(x; F_1)$ with $F_1$ denotes the first marginal distribution of $F$.
Let $\Phi$ and $\widetilde\Phi$ be two grading matrices such that $\Phi(1,\cdot)=\widetilde\Phi(1,\cdot)$. Then the empirical distribution associated respectively to $\Phi$ and $\widetilde\Phi$ have the same first marginal $F_{\Phi_1}$. Denote respectively $x^*$ and $\widetilde x^*$ two deepest points for $\Phi$ and $\widetilde\Phi$. Then both $x^*_1$ and $\widetilde x^*_1$ belong to $\argmax_{x\in\R} D_1(x; \Phi(1,\cdot))$. \ref{C5} is hence satisfied.

Now, consider the converse implication. 
Let 
\begin{eqnarray*}
G_1:&~ \R^{d\times n} &\to \mathcal B(\R)\\
& \Phi & \mapsto \mathcal X^*_1=\{x_1^*, \; x^*\in\argsup_{x\in \mathbb{R}^d} D(x; \Phi)\}.
\end{eqnarray*} 
If $G_D$ satisfies \ref{C5}, then for all $\Phi=(\Phi(\cdot,j))_{j=1,\dots,n}\in\R^{d\times n}$, $G_1(\Phi(1,\cdot),\dots,\Phi(d,\cdot))$ is a constant set $\mathcal G_1(\Phi(1,\cdot))$. Define $D_1$ as 
\begin{eqnarray*}
 D_1: &~ \R\times\R^{n} &\to \R\\
& (x,\Phi(1,\cdot))  &\mapsto \begin{cases}
1 & \text{if } x\in \mathcal G_1(\Phi(1,\cdot))\\
0 & \text{else}
\end{cases}.
\end{eqnarray*} 
Then for all $\Phi\in\R^{d\times n}$, for all $x^*\in \argsup_{x\in\R^d}  D(x; \Phi)$, the first component satisfies $x_1^*\in \argsup_{x\in\R} D_1(x, \Phi(1,\cdot))$.

%%%%%%%%%%%%%%%%%%%%%%%%%%%%%%%%%%%%%%%%%%%%%%%%%

\subsection{Proof of Proposition \ref{prop:C4sansC5}}

Let $D$ be the depth function defined in \eqref{eqn:defZ}. For any grading matrix $\Phi$, for any level $a\in(0,1)$, define the contour sets of the zonoïd depth, $D_{Za}(\Phi)=\{x\in\R^d,~D_Z(x;\Phi)\geq a\}.$ The deepest point of the depth function defined in \eqref{eqn:defZ} is the coordinate-wise mean of the points in the set $D_{Z 0.8}$.

Let us first prove that properties \ref{ass:depth0} to \ref{ass:depth5} hold.
\ref{ass:depth0}, \ref{ass:depth4} and \ref{ass:depth5} are straightforward, since they are satisfied by the zonoïd depth function. \ref{ass:depth1} follows from the fact that it is satisfied by both the zonoïd depth function and the mean operator.

Let us study \ref{ass:depth3}. Let $\theta$ be the deepest point for a distribution $F$. 
\begin{itemize}
\item Let $x\in D_{Z0.8}(F)\setminus\{\theta\}$. By construction, $D(x)=0.8$. For all $0< \lambda \leq 1$, $\theta+\lambda (x-\theta)\in  D_{Z0.8}(F)$, since the contour sets of the zonoïd depth are convex (see \cite[page 412]{WeightedMeans}). Hence $D(x)=D(\theta+\lambda (x-\theta))=0.8$. 
\item Now consider $x\notin D_{Z0.8}(F)$. The contour sets of the zonoïd depth are nested  (see \cite[Proposition 3]{WeightedMeans}). Hence, for all $0\leq \lambda\leq 1$, $\theta+\lambda (x-\theta)\in D_{Z a}(F)$, with $a=D_Z(x)$.  
\end{itemize}
Hence, \ref{ass:depth3} holds.

Let us study \ref{ass:depth2}. Weighted mean depths satisfy property \ref{ass:depth2} with respect to central symmetry. A set $S \subset \R^d$ is centrally symmetric with center $c \in\R^d$, if for every point $c + d \in S$ the point $c - d$ is also in $S$. Let $F$ be a distribution which is centrally symmetric about some $c \in\R^d$. As stated by \cite[Corollary 1]{WeightedMeans}, then $c\in D_{Z0.8}$. And for every point $c + d \in D_{Z0.8}$ the point $c - d$ is also in $D_{Z0.8}$. Hence, the %arithmetic
 mean of $D_{Z0.8}$ coincides with $c$. It concludes the proof.

Since \ref{ass:depth0} to \ref{ass:depth5} hold, the deepest voting procedure satisfies \ref{C1}, \ref{C2}, \ref{C3} by Proposition \ref{prop:lien}.

Now consider \ref{C4}. Since the zonoïd depth belongs to the family of weighted mean depths, it is monotone in the data \citep{WeightedMeans}. Let us explicit this property. Let $\Phi$ and $\widetilde \Phi$ be two grading matrices such that $\Phi = \widetilde\Phi$ except that one or more voters give higher grades to a candidate in $\Phi $ than in $ \widetilde\Phi $. Then, for all $a\in(0,1)$,
\begin{equation*}
D_{Za}(x;\Phi) \subseteq D_{Za}(x;\widetilde\Phi)\}\oplus \R_+^d
\end{equation*}
where $\oplus$ denotes the Minkowski sum of sets. See \cite[Proposition 8]{WeightedMeans}. 
Consequently, taking $a=0.8$,  $D_{Z0.8}(x;\Phi) \subseteq D_{Z0.8}(x;\widetilde\Phi)\}\oplus \R_+^d$. It is easily seen that the deepest point for $\widetilde\Phi$ has higher coordinates than the deepest point for $\Phi$. This proves \ref{C4}.

\begin{figure}[!ht]
\centering
\includegraphics[height=6cm]{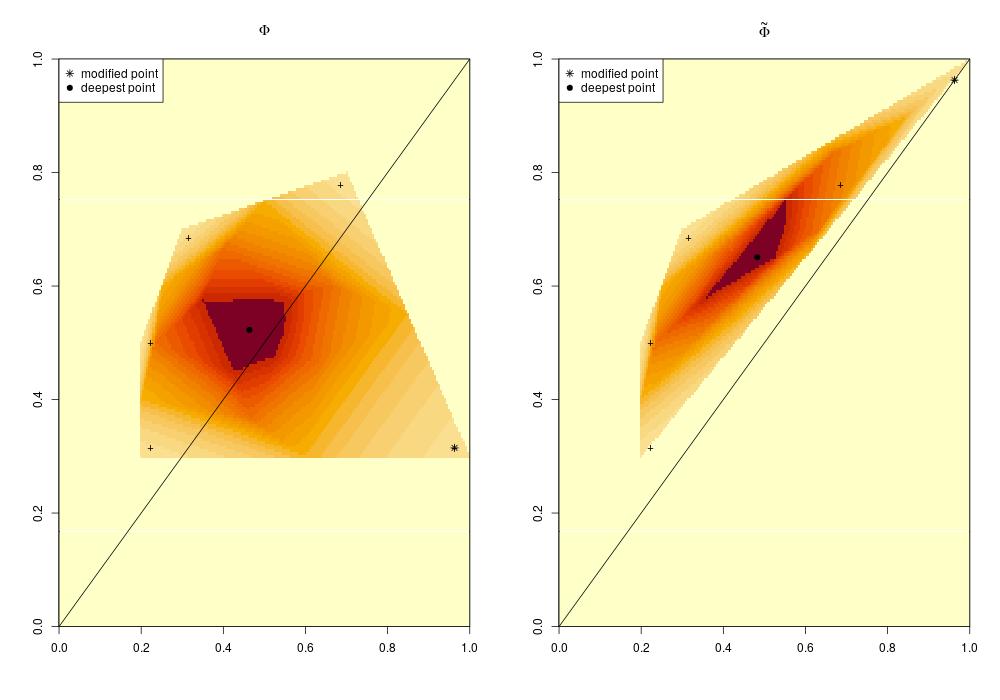}
\caption{ Deepest points based on the depth function defined in \eqref{eqn:defZ} for grading matrices defined in \eqref{eqn:IIAzonoid}. Horizontal axes give the grade for candidate $c_1$ and vertical axes for candidate $c_2$. Each cross corresponds to a voter.}
\label{fig:zonoid}
\end{figure}

What is left is to show that \ref{C5} does not hold. It is sufficient to provide a counterexample.
Let
\begin{align} \label{eqn:IIAzonoid}
\Phi&=\begin{pmatrix}
 1.0 & 0.2 & 0.2 & 0.7 & 0.3\\
 0.3 & 0.5 & 0.3 & 0.8 & 0.7
\end{pmatrix},\\ \nonumber
\widetilde\Phi&=\begin{pmatrix}
1.0 & 0.2 & 0.2 & 0.7 & 0.3\\
1.0 & 0.5 & 0.3 & 0.8 & 0.7
 \end{pmatrix}.
\end{align}
The numerical application gives that the deepest point for $\Phi$ is $x^*=(0.46, 0.52)$ when the deepest point for $\widetilde\Phi$ is $\widetilde x^*=(0.48, 0.66)$. See Figure~\ref{fig:zonoid}. The first coordinate of the deepest point has changed, while only the votes for the second candidate have changed. Hence \ref{C5} is not satisfied.

%%%%%%%%%%%%%%%%%%%%%%%%%%%%%%%%%%%%%%%%%%%%%%%%%

\subsection*{Proof of Table \ref{tab:depths}}

The unicity of the deepest point for $L^p$ depths is proved in Proposition~\ref{p2} when $1<p<\infty$ and in Proposition~\ref{p3} when $p=\infty$. IIA assumption is straightforward with Proposition~\ref{prop:IIA}, since maximizing the $L^p$ depth is equivalent to minimizing $x\mapsto\sum_{i=1}^d\sum_{j=1}^n (x-\Phi(i,j))^p$.

To establish that \ref{C5} is not satisfied, let us provide a counter-example. Observe that a counter-example has already been given in page \pageref{ex:iia} for the halfspace depth. We provide another counter-example to prove the result for the Oja depth and the projection depth.

Let
\begin{align*}
\Phi&=\begin{pmatrix}
 0.4 & 0.6 & 0 & 0.4 & 0.6 & 0 & 0.3\\
 0.7 & 0 & 0.6 & 0.4 & 0.4 & 0.8 & 0
\end{pmatrix},\\
\widetilde\Phi&=\begin{pmatrix}
 0.4 & 0.6 & 0 & 0.4 & 0.6 & 0 & 0.3\\
 0.7 & 0 & 0.6 & 0.4 & 0.4 & 0.8 & 1
 \end{pmatrix}.
\end{align*}
The numerical application gives the following deepest points for $\Phi$,

\begin{center}
\begin{tabular}{lcc}
\hline
& \multicolumn{2}{c}{$\Phi$}  \\
Depths  & deepest grade for $c_1$ & deepest grade for $c_2$ \\ \hline
Halfspace depth & 0.38  & 0.40 \\
Projection depth  & 0.37 & 0.43 \\
Oja depth        & 0.40  & 0.40 \\ \hline
\end{tabular}
\end{center}

and the following deepest points for $\widetilde\Phi$,

\begin{center}
\begin{tabular}{lcc}
\hline\\ [-10pt]
& \multicolumn{2}{c}{$\widetilde\Phi$}  \\
Depths  & deepest grade for $c_1$ & deepest grade for $c_2$ \\ \hline
Halfspace depth & 0.34 & 0.59 \\
Projection depth & 0.32  & 0.57 \\
Oja depth       & 0.34  &  0.59   \\ \hline
\end{tabular}
\end{center}

It can be seen that the change of one vote for candidate $c_2$ influences the result for candidate $c_1$. Hence, by Proposition~\ref{prop:IIA}, we conclude that \ref{C5} is not satisfied.

\cite[Proposition 5]{WeightedMeans} establish that the deepest point of weighted mean depths is the component-wise mean of the observations. It is, hence, unique and satisfies the IIA property.

%%%%%%%%%%%%%%%%%%%%%%%%%%%%%%%%%%%%%%%%%%%%%%%%%%%%%%%%%%%%%%%%%%%%%%%%%%%%%%%%%%%%%%%%%%

\subsection{Proof of Proposition \ref{p1bis}}

As stated above, each coordinate of a deepest point $x^*:=(x^*_1,\hdots,x^*_d)$ is given by the optimization problem:
\begin{equation*}
\forall i = 1,\hdots,d, \quad x^*_i=\argmin_{x\in \mathbb{R}}  \sum_{j=1}^n |\Phi(i,j)-x|^p.
\end{equation*}
Since the function $x\mapsto x^p$ is strictly convex for all $p>1$, it results that $x_i^*$ exists and is unique for all $i=1,\dots,d$.

%%%%%%%%%%%%%%%%%%%%%%%%%%%%%%%%%%%%%%%%%%%%%%%%%%%%%%%%%%%%%%%%%%%%%%%%%%%%%%%%%%%%%%%%%%
\subsection{ Proof of Proposition \ref{p2}}

The proof is trivial since it is well-known that the quantity $x$ minimizing 
$\sum_{j=1}^n |\Phi(i_0,j)-x|^2$ (resp. $\sum_{j=1}^n |\Phi(i_0,j)-x|$) is the sample mean 
(resp. the median) of $(\Phi(i_0,1), \hdots, \Phi(i_0,n)).$
The same result holds for approval voting, since it corresponds to range voting when the possible grades are restricted to 0 and 1.

%%%%%%%%%%%%%%%%%%%%%%%%%%%%%%%%%%%%%%%%%%%%%%%%%%%%%%%%%%%%%%%%%%%%%%%%%%%%%%%%%%%%%%%%%%%

\subsection{ Proof of Proposition \ref{p3}}

Consider the ordered grades $r_1 \leq r_2 \leq \hdots \leq r_n$ of a given candidate $c_i$, $i=1,\dots,d$.
We want to prove that the point $x_0=\frac{r_1+r_n}{2}$ minimizes the function  $g: x \mapsto \max_{j=1,\dots,n} |r_j-x|$.
 Remark that as we ordered the values, $g(x) = \max \{|r_1-x|,|r_n-x|\}$, and thus
\[
g(x) = \begin{cases}
r_n -x> r_n-x_0& \text{ if } x<x_0\\
x_0-r_1=r_n-x_0& \text{ if } x=x_0\\
x-r_1 > x_0-r_1& \text{ if } x>x_0\,.
\end{cases}
\]

%%%%%%%%%%%%%%%%%%%%%%%%%%%%%%%%%%%%%%%%%%%%%%%%%%%%%%%%%%%%%%%%%%%%%%%%%%%%%%%%%%%%%%%%%%
\subsection{ Proof of Proposition \ref{p4}}

We present configurations such that 
\begin{itemize}
\item a Condorcet winner is not elected by $L^p$ Deepest Voting,
\item a Condorcet loser is elected by $L^p$ Deepest Voting.
\end{itemize}
We distinguish the two cases $p=1$ and $p>1$.

\begin{itemize}
\item  Case p=1.

Consider the following configuration with 9 voters and 3 candidates:
\begin{center}
\begin{tabular}{cccc}
\hline 
Number of voters & Grade for $c_1$ &  Grade for $c_2$ & Grade for $c_3$\\ 
\hline 
4 &  0.5 & 0.1 & 0.4 \\
1 & 0.5 & 0.6 & 0.4 \\
4 & 1 & 0.6 & 0.7 \\
\hline 
\end{tabular} 
\end{center}
The representation of the votes in the space of candidate $c_1$ and candidate $c_2$ is given in Figure \ref{NI1} taking $(a,b,c,d,p_0)=(0.1,0.5,0.6,1,1/9)$. 
As the median of grades for candidate $c_2$ is higher than the median for the other candidates, candidate $c_2$ is elected by $L^1$ deepest voting. However, $c_1$ is the Condorcet winner since an absolute majority of voters prefer $c_1$ to $c_2$ and $c_3$. In the same time, $c_2$ is elected even if he is the Condorcet loser, as $c_1$ and $c_3$ are preferred to $c_2$ by a majority of voters.

\begin{figure}[!ht]
\centering
\includegraphics[width=0.75\textwidth]{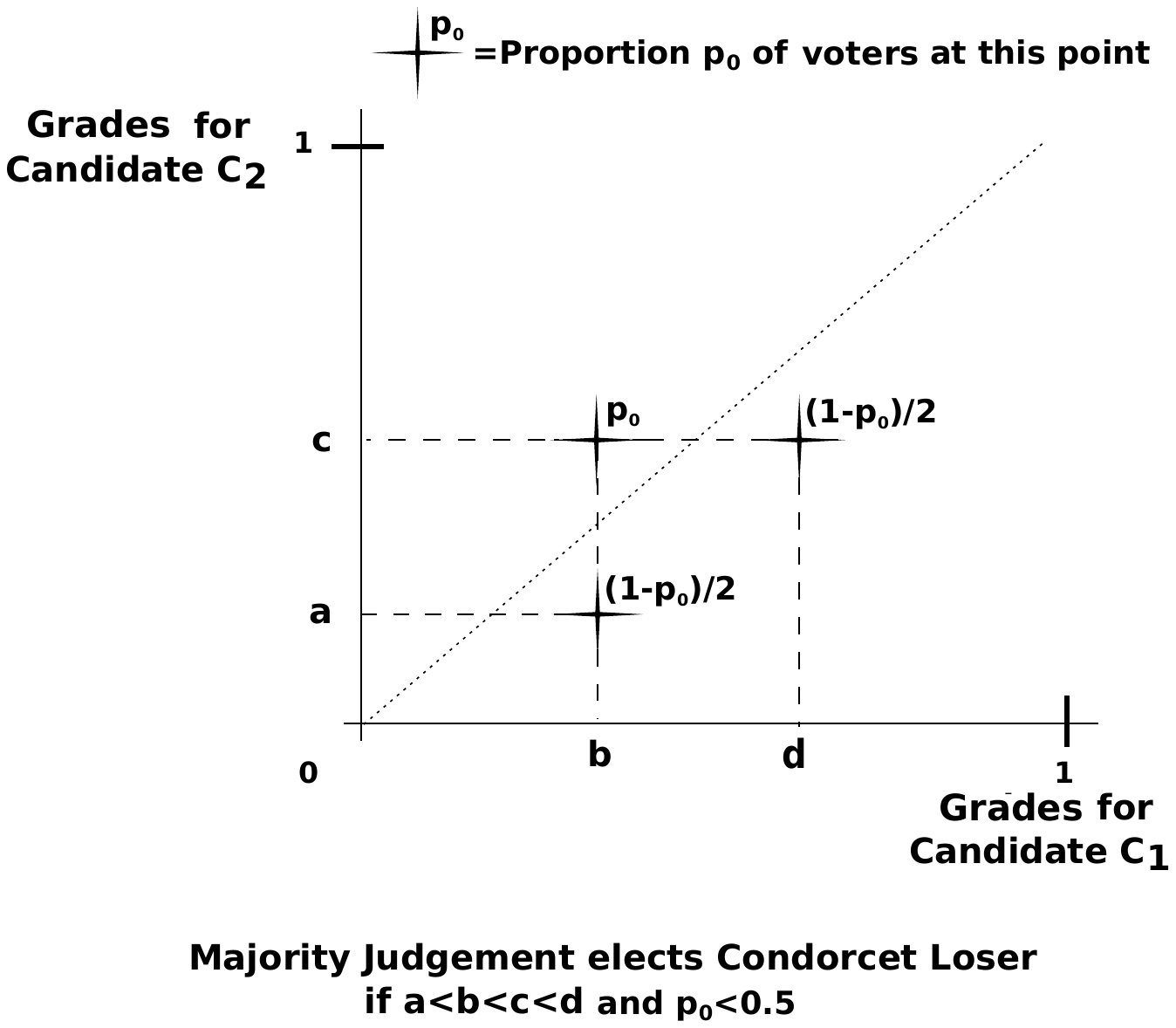}
\caption{ Configuration of grades considered in the proof of Proposition~\ref{p4} in the case $p=1$. }
\label{NI1}
\end{figure}

\item Case $p>1$. 

Consider the following profiles for $3$ candidates and $n$ voters:
\begin{center}
\begin{tabular}{cccc}
\hline 
Number of voters & Grade for $c_1$ &  Grade for $c_2$ &  Grade for $c_3$ \\ 
\hline 
$n-1$ &  $0.5 + \varepsilon_{n,p}$ & 0.5 & $0.5 + \varepsilon_{n,p}/2$\\
1 & 0 & 1 & 0\\
\hline 
\end{tabular} 
\end{center}
with $0<\varepsilon_{n,p} \leq \min(0.5 ; (n-1)^{\frac{-1}{p-1}})$.
The representation of the votes in the space of candidate $c_1$ and candidate $c_2$ is given in Figure \ref{NI2}, taking $p_0=1/n$.

\begin{figure}[!ht]
\includegraphics[width=0.75\textwidth]{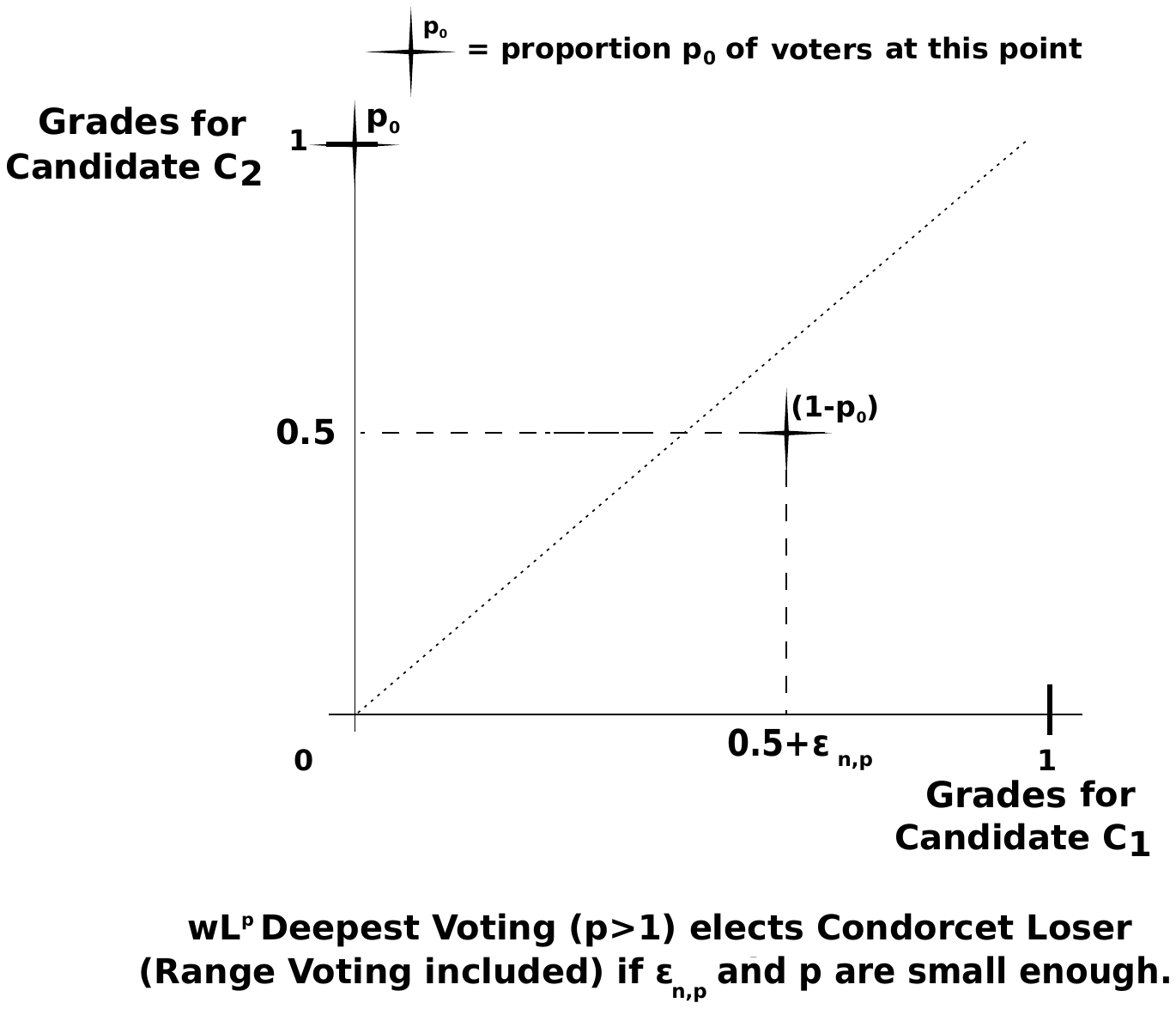}
\caption{Configuration of grades considered in the proof of Proposition~\ref{p4} in the case $p>1$. }
\label{NI2}
\end{figure}

The coordinates of the deepest point $x^*_p=(x^*_{p,1},x^*_{p,2}, x^*_{p,3})$ are solutions of the following optimization problems:
\begin{align*}
x^*_{p,1}&= \arg \displaystyle \min_{x\in [0,1]} (n-1) |x-(0.5+\varepsilon_{n,p})|^p + x^p \\
x^*_{p,2}&= \arg \displaystyle \min_{x\in [0,1]} (n-1) |x-0.5|^p + |1-x|^p\\
x^*_{p,3}&= \arg \displaystyle \min_{x\in [0,1]} (n-1) |x-(0.5+\varepsilon_{n,p}/2)|^p + x^p .
\end{align*}
Using the first order conditions of the optimization problems, we get:
\begin{align*}
x^*_{p,1}&= \frac{(0.5+\varepsilon_{n,p})(n-1)^{\frac{1}{p-1}}}{(n-1)^{\frac{1}{p-1}}+1}\\
x^*_{p,2}&= \frac{1+0.5(n-1)^{\frac{1}{p-1}}}{1+(n-1)^{\frac{1}{p-1}}}\\
x^*_{p,3}&= \frac{(0.5+\varepsilon_{n,p}/2)(n-1)^{\frac{1}{p-1}}}{(n-1)^{\frac{1}{p-1}}+1}.
\end{align*}
When $0<\varepsilon_{n,p} \leq (n-1)^{\frac{-1}{p-1}}$,it follows that $ x^*_{p,3}<x^*_{p,1} < x^*_{p,2}$.  Consequently, candidate $c_2$ (who is the Condorcet loser) is elected by  $L^p$-deepest voting even if the Condorcet winner is $c_1$.
\end{itemize}

%%%%%%%%%%%%%%%%%%%%%%%%%%%%%%%%%%%%%%%%%%%%%%%%%%%%%%%%%%%%%%%%%%%%%%%%%%%%%%%%%%%%%%%%%%
\subsection{Proof of Proposition \ref{p5}}

We distinguish according to the values of $p$.
\begin{itemize}

\item Case $p=1$ 

For an example proving the vulnerability of the majority judgment to reinforcement (resp. no-show) paradox, see \cite[p. 327 (resp. p. 329)]{Felsenthal}.

\item Case $1<p<2$ 

Consider the following configuration of grades:
\begin{center}
\begin{tabular}{lcc} \hline
         & Grade for $c_1$  & Grade for $c_2$\\ \hline
Voter $v_1$  & 0.5 & 1\\
Voter $v_2$  & 0.5 & $\varepsilon$\\
Voter $v_3$  & 0 & $\varepsilon$\\ \hline
\end{tabular}
\end{center}
with $0<\varepsilon<1$.
If only voters $v_1$ and $v_2$ are voting, then candidate $c_2$ wins, whatever $\varepsilon>0$. Moreover, voter $v_3$ prefers candidate $c_2$. Hence, to prove that no-show and reinforcement paradoxes hold, we may establish that considering voters $\{v_1, v_2, v_3\}$, candidate $c_2$ may loose.

Denote $x^*:=(x_1^*,x_2^*)$ the deepest point obtained for the $L^p$ depth with voters $(v_1, v_2, v_3)$.

We aim at proving that $x_2^*<x_1^*$ for some $0<\varepsilon<0.5$.

For $i=1, 2$, components of the deepest point  $x^*$ are obtained by 
 \[
x_i^* = \argmin_{x\in \mathbb{R}} \sum_{j=1}^n \lvert \Phi(i,j) - x \rvert^p.
\]
Differentiating the objective function with respect to $x$ leads to 
\[
\sum_{j=1}^n \lvert\Phi(i,j) - x \rvert^{p-1}\text{sgn}(\Phi(i,j) - x).
\]
First order conditions for $i=1$ lead to
\[
2\lvert x_1^*-0.5\rvert^{p-1}\,\text{sgn}(x_1^*-0.5) + \lvert x_1^*\rvert^{p-1}\,\text{sgn}(x_1^*)=0
\]
Consequently, $0< x^*_1 <0.5$ and we deduce that $x_1^*=0.5\,\frac{2^{1/(p-1)}}{1+2^{1/(p-1)}}$.

For $i=2$ we get
\[ 
2\lvert\varepsilon- x_2^*\rvert^{p-1}\,\text{sgn}(\varepsilon-x_2^*) + \lvert 1-x_2^*\rvert^{p-1}\,\text{sgn}(1-x_2^*)=0
\]
Similarly, we deduce that $x_2^*=\frac{1+2^{1/(p-1)}\varepsilon}{1+2^{1/(p-1)}}$.

Thus
\[
x_2^\ast<x_1^\ast \iff  \varepsilon< 0.5-{2^{-1/(p-1)}}
\]
Since the right-hand side is positive for $p<2$, we deduce that there exists $\varepsilon>0$ such that $x_2^\ast<x_1^\ast$. This concludes the proof.

\item Case $p=2$

$L^2$ deepest point is the point with coordinates equal to the mean of votes. Let $\overline{x_1}$ and $\overline{y_1}$ be the mean score obtained respectively by two candidate $c_1$ and $c_2$ on a population of voters of size $n_1$ and let $\overline{x_2}$ and $\overline{y_2}$ be the mean score obtained respectively by the two candidates $c_1$ and $c_2$ on a population of size $n_2$. If $\overline{x_1}<\overline{y_1}$ and $\overline{x_2}<\overline{y_2}$, then $n_1\overline{x_1}+n_2\overline{x_2}< n_1\overline{y_1}+n_2\overline{y_2}$. Hence, if candidate $c_2$ wins in the two sub-populations, he also wins on the total population. We can deduce that reinforcement paradox does not hold. Next, taking $n_2=1$, it implies that no-show paradox does not hold either.

\item Case $2<p$

For $0<\varepsilon<0.5$, consider the following grades:
\begin{center}
\begin{tabular}{lcc} \hline
         & Grade for $c_1$  & Grade for $c_2$\\ \hline
voter $v_1$  & 0 & 0.5+$\varepsilon$\\
voter $v_2$  & 1 & 0.5+$\varepsilon$\\
voter $v_3$  & 0 & $\varepsilon$\\ \hline
\end{tabular}
\end{center}
Then if only voters $v_1$ and $v_2$ are voting, candidate $c_2$ wins because $0.5+\varepsilon > 0.5$. Since $v_3$ prefers $c_2$, to prove that no-show and reinforcement paradoxes hold, let's establish that considering voters $\{v_1, v_2, v_3\}$, candidate $c_2$ may loose.

Denote $x^*:=(x_1^*,x_2^*)$ the deepest point obtained for the $L^p$ depth with voters $(v_1, v_2, v_3)$.

We aim at proving that $x_2^*<x_1^*$ for some $0<\varepsilon<0.5$.

For $i=1, 2$, components of the deepest point  $x^*$ are obtained by 
 \[
x_i^* = \argmin_{x\in\R} \sum_{j=1}^n \lvert \Phi(i,j) - x \rvert^p.
\]
First order conditions for $i=1$ imply
\[
2\,\lvert x_1^*\rvert^{p-1}\,\text{sgn}(x_1^*) + \lvert 1-x_1^*\rvert^{p-1}\,\text{sgn}(1 - x_1^*)=0.
\]
Hence $0 < x^*_1 <1$ and we obtain that $x_1^*=\frac{1}{1+2^{1/(p-1)}}$. 

First order conditions for $i=2$ lead to
\[
2\,\lvert 0.5+\varepsilon-x_2^*\rvert^{p-1}\,\text{sgn}( 0.5+\varepsilon-x_2^*) + \lvert \varepsilon-x_2^*\rvert^{p-1}\,\text{sgn}(\varepsilon - x_2^*)=0.
\]
We deduce analogously that $x_2^*=0.5\,\frac{2^{1/(p-1)}}{1+2^{1/(p-1)}}+\varepsilon$. 

Thus
\begin{align*}
x_2^\ast<x_1^\ast & \iff  \varepsilon< \frac{1}{1+2^{1/(p-1)}}-\frac{1}{2}\frac{2^{1/(p-1)}}{1+2^{1/(p-1)}}\\
&\iff \varepsilon < \frac{2-2^{1/(p-1)}}{2\,(1+2^{1/(p-1)})}.
\end{align*}
Since the right-hand side is positive for $p>2$, we deduce that there exists $\varepsilon>0$ such that $x_2^\ast<x_1^\ast$. This concludes the proof.

\end{itemize}
%%%%%%%%%%%%%%%%%%%%%%%%%%%%%%%%%%%%%%%%%%%%%%%%%%%%%%%%%%%%%%%%%%%%%%%%%%%%%%%%%%%%%%%%%%
\newpage

\bibliographystyle{abbrvnat}
\bibliography{bibvote}

%%%%%%%%%%%%%%%%%%%%%%%%%%%%%%%%%%%%%%%%%%%%%%%%%%%%%%%%%%%%%%%%%%%%%%%%%%%%%%%%%%%%%%%%%%

%%%%%%%%%%%%%%%%%%%%%%%%%%%%%%%%%%%%%%%%%%%%%%%%%%%%%%%%%%%%%%%%%%%%%%%%%%%%%%%%%%%%%%%%%%
\end{document}